# Cycles, determinism and persistence
# in agent-based games and financial time-series

J.B. Satinover[1] and D. Sornette[2]

[1]*Laboratoire de Physique de la Matière Condensée*
*CNRS UMR6622  and Université des Sciences, Parc Valrose*
*06108 Nice Cedex 2, France*
[2]*Department of Management, Technology and Economics*
*ETH Zurich, CH-8032 Zurich, Switzerland*
jsatinov@princeton.edu and dsornette@ethz.ch

May 3, 2008

The Minority Game (MG), the Majority Game (MAJG) and the Dollar Game ($G) are important and closely-related versions of market-entry games designed to model different features of real-world financial markets. In a variant of these games, agents measure the performance of their available strategies over a fixed-length rolling window of prior time-steps. These are the so-called Time Horizon MG/MAJG/$G (THMG, THMAJG, TH$G)s. Their probabilistic dynamics may be completely characterized in Markov-chain formulation. Games of both the standard and TH variants generate time-series that may be understood as arising from a stochastically perturbed determinism because a coin toss is used to break ties. The average over the binomially-distributed coin-tosses yields the underlying determinism. In order to quantify the degree of this determinism and of higher-order perturbations, we decompose the sign of the time-series they generate (analogous to a market price time series) into a superposition of weighted Hamiltonian cycles on graphs—exactly in the TH variants and approximately in the standard versions. The cycle decomposition also provides a "dissection" of the internal dynamics of the games and a quantitative measure of the degree of determinism. We discuss how the outperformance of strategies relative to agents in the THMG—the "illusion of control"—and the reverse in the THMAJG and TH$G, i.e., genuine control—may be understood on a cycle-by-cycle basis. The decomposition offers as well a new metric for comparing different game dynamics to real-world financial time-series and a method for generating predictors. We apply the cycle predictor a real-world market, with significantly positive returns for the latter.

## 1. OVERVIEW

The Minority Game (MG), the Majority Game (MAJG) and the Dollar Game ($G) are multi-agent games in which agents make binary decisions in an attempt to maximize their individual gain at each time-step. Their decisions represent a response to the summed state of all agents' prior decisions. In the MG agents attempt to make the same decision as a minority of all other agents. In the MAJG and $G agents attempt to make the same decision as the majority of all other agents. (The distinction between MAJG and $G is in how agents' gain is computed as detailed below.)

The MG, MAJG and \$G are all pure versions of market-entry games. They have attracted keen interest in the physics, economics and optimization communities—especially the MG because of its intrinsic frustration, multiple equilibria and similarity to multi-body spin systems.

When past information is limited to a rolling window of prior states of fixed length $\tau$, the MG, MAJG and \$G may all be expressed in Markov-chain formulation [1-3]. (These are the "time-horizon" variants of the games: THMG, THMAJG, TH\$G.) If a standard game reaches its steady-state in $\tau'$ steps, its dynamics are equivalent to its TH variant with window length $\tau'$.

Here, we present a new use of a cycle decomposition method that expresses the inherently probabilistic nature of a Markov chain as an exact superposition of deterministic sequences, extending ideas discussed in [4]. The nature of the underlying series and/or deterministic cycles is related in complex fashion to the degree of "persistence" or of "anti-persistence" they display: "Persistence" is a quantitative measure of the tendency of patterns in a time-series to be followed by repetitions of that same pattern. Similarly, "anti-persistence" is a measure of the tendency of patterns to be followed rather by their respective inverses. (Exact definitions and procedures for obtaining these measures follow). On the unit interval perfect persistence $\mathcal{P}$ may be defined as "1" and perfect anti-persistence as "0" with random sequences having $\mathcal{P} = 0.5$.

Ultimately, these games are meant to illuminate the behavior of real-world financial markets and the time-series they generate. Analyses of both cycles and persistence shed light on real-life financial time series, in particular on their departures from randomness. An important question that has been addressed sporadically is whether the light shed by studies of these games may be translated into methods applicable to, for example, the prediction of real-world time-series. We provide an example of how insights derived from these games may indeed be translated into real-world prediction methods. We do so by constructing a predictor based on a decomposition of the time-series into weighted deterministic cycles on graphs and by quantifying the degree of persistence or anti-persistence in our target series. For pedagogical purposes we here demonstrate only "toy" predictors of great simplicity. However, these methods may be readily generalized and improved, opening up new avenues for research and application, for instance, by encoding price-change histories in trinary or even more detailed form. The cycle decomposition method for such higher-order matrices is significantly more complex, however.

## 2. SUMMARY OF FINDINGS

In this paper we present six findings and/or related applications:

First, as presented in full detail in [3], we note that for the two games employing the majority rule (MAJG, $G), A(t) (the binary time series generated by agents in the aggregate) is persistent ($\vartheta > 0.5$), approaching the random limit asymptotically with increasing memory length $m$ (i.e., $\lim_{m \to \infty} \vartheta = 0.5$); in minority games the time series crosses from anti-persistent for $m < m_c$ ($\vartheta < 0.5$) to persistent ($\vartheta > 0.5$) at the well-known phase transition, attains a maximum and then declines asymptotically to ($\vartheta = 0.5$) with further increasing $m$.

Second, we demonstrate how the time series generated by the THMG, THMAJG and TH$G (all of which are Markovian processes) may be exactly represented as a weighted superposition of deterministic cycles on graphs.

Third, a decomposition of the respective series into such cycles on graphs reveals in highly intuitive fashion characteristic differences among the three types of games. These differences in cycle structure are consistent with, but further differentiate, the distinction between persistent and anti-persistent series.

Fourth, we apply the cycle decomposition method to the prediction of game-generated time-series, based on a sliding window of past information.

Fifth, all three types of game-generated time-series as well as real world series may be reformulated as perturbations of a characteristic underlying dynamic that to the zero[th] order is wholly deterministic. I.e., any binary time series may be decomposed into a superposition of wholly deterministic Hamiltonian cycles on graphs. The cycle of greatest weight may be considered the dominant underlying determinism; cycles of lesser weight may be thought of as the higher-order perturbations. Probabilistic transitions from one history to another are in this view recast as probabilistic transitions among cycles. This cycle decomposition approach parallels the theories of dynamics systems and of deterministic chaos [5] in particular on the one hand and of quantum chaos [6] on the other hand, both based on the decomposition on unstable periodic orbits.

Sixth is the application of the cycle decomposition method and an analysis of persistence to real-life financial time series: Different series (or the same series in different periods) may be characterized by a signature cycle structure and/or degree of persistence. This fact leads to prediction methods based on cycles and on persistence, even for time-series which are not in fact Markovian.

## 2.1 PERSISTENCE VERSUS ANTI-PERSISTENCE IN THE MG, MAJG AND $G

Within a long binary string, consider a possible sub-sequence of length $m'$. Examining the string from one end to the other, compute the number of times that, when the sub-sequence is followed by the digit 1, the *next* appearance of the sub-sequence is followed by the digit 0. Tally this as one instance of "anti-persistence". (If the sub-sequence is again followed by 1, tally one instance of "persistence"). Re-examine and tally the string in the same way for all possible subsequences of length $m$. Then the overall proportion of persistent tallies is the strings' "persistence" at scale $m$; 1 minus its persistence is the series' anti-persistence at scale $m$.

Among other features to be discussed later and previously discussed in detail in [3] **Figure 1** illustrates that in terms of degree of persistence $\mathcal{P}$, $\mathcal{P}_{min} < \mathcal{P}_{S} < \mathcal{P}_{maj}$. This holds true for all memory lengths $m'$ in the respective games and (to be explained later, as also in ref. [3]) at the natural scale at which the persistence is measured. (I.e., the set of patterns being examined in a series for repetition is determined by a bit-length $m'$, and the size of that set grows as $m'^2$. It is both convenient and natural to set $m' = m$ when analyzing persistence in the MG, MAJG and $G, where $m$ is the known bit-length of the history accessible to the games' agents at each time-step). **Figure 1** also illustrates that the variability in persistence differs from game to game and for the MG by phase.

Consistent with the fact that for $m < m_c$ the time-series itself is highly variable from run to run and much less so for $m \geq m_c$, the variability in persistence from run to run likewise changes at the phase transition. Consistent with it high degree of persistence, there is little variability altogether for the MAJG. The variability in $\mathcal{P}$ for the $G is roughly consistent throughout and much greater than for the MAJG.

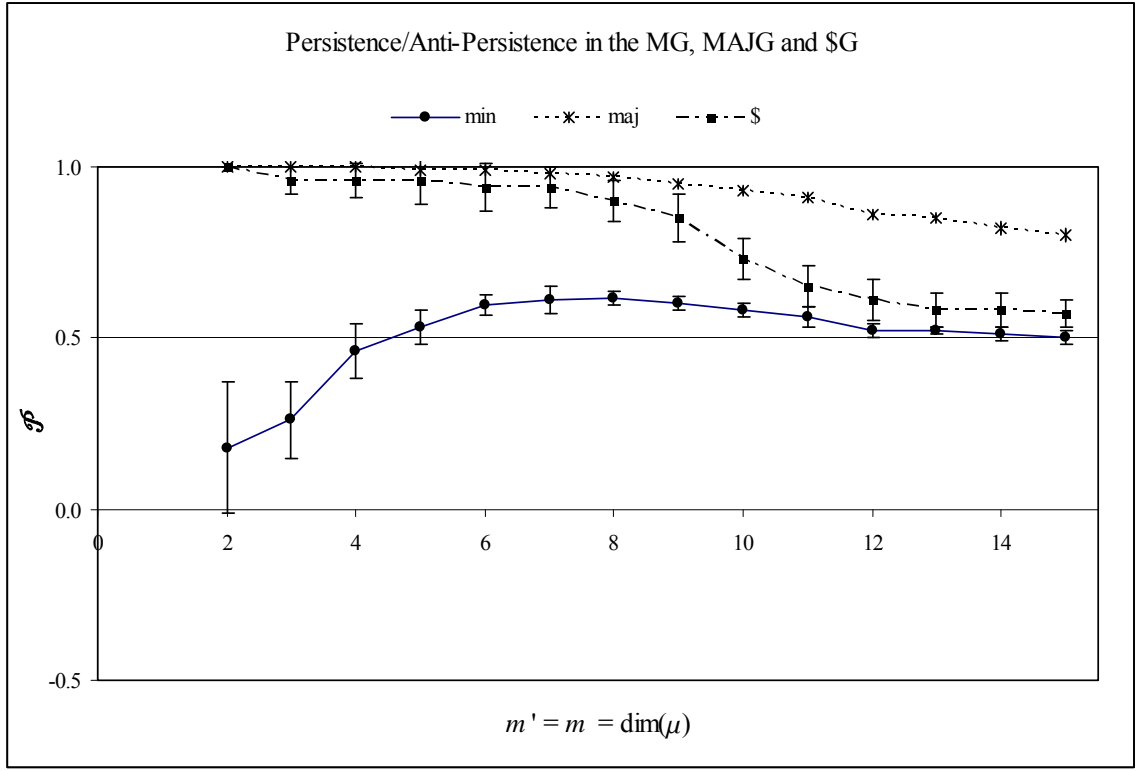

**Figure 1**: Persistence and anti-persistence in MG, MAJG and $G for $m = m'$, 100 runs of 1000 time steps for each game, at each $m' = m$ Error bars show 1 SD (barely visible at this scale for the MAJG).

## 2.2 CYCLE STRUCTURE OF THE MG, MAJG AND $G

We may also compare MG, MAJG and $G in terms of the decomposition of the binary time-series they generate into Hamiltonian cycles on graphs (at some given length $m'$). To do this, we first encode a binary series, i.e. as its decimal equivalent +1 given length $m'$ (a rolling window). For example, at $m'=2$:

$$\{0,1,1,1,0,0,1,0,\ldots\} \rightarrow \{(0,1),(1,1),(1,1),(1,0),(0,0),(0,1),(1,0)\ldots\}$$
$$\rightarrow \{1,3,3,2,0,1,2\ldots\} + 1 = \{2,4,4,3,1,2,3\ldots\}$$

(1)

The allowed transitions from one $m'$-bit state to the next form a complete binary de Bruijn graph of order $m'=2$ as shown in the middle graphic of **Figure 2**. (examples of complete de Bruijn graphs of order $m'=1$ and $m'=3$ are shown to the left and right respectively).

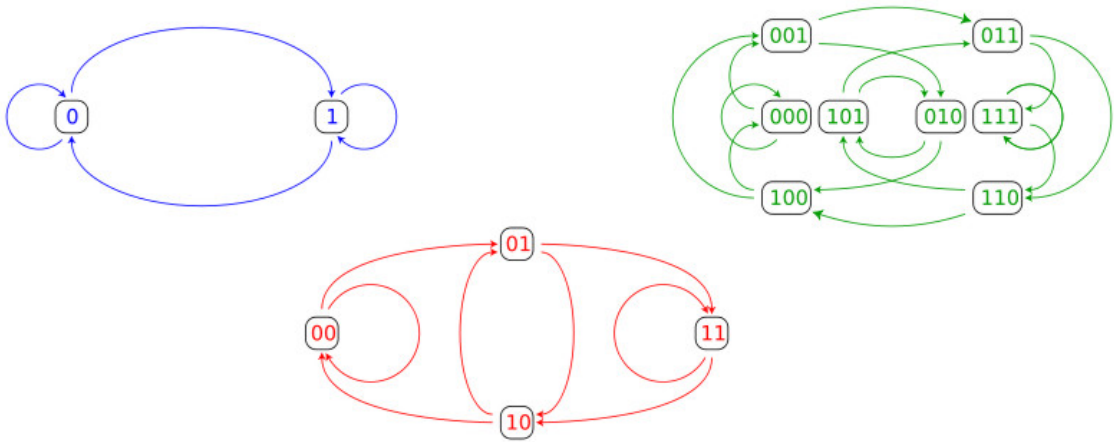

**Figure 2**: Complete binary de Bruijn graphs of orders 1, 2 and 3 from left to right. The vertices may be numbered as shown or by their decimal equivalents +1. In a complete graph, all possible states and transitions are represented.

If a binary sequence thus encoded touches no vertex more than once except upon returning to the first, the sequence is considered a cycle. **Figure 3** shows the cycle consisting of the last four digits in (1). The four digits are represented as three vertices (and the "edges" connecting them) because the last digit (vertex) repeats the first.

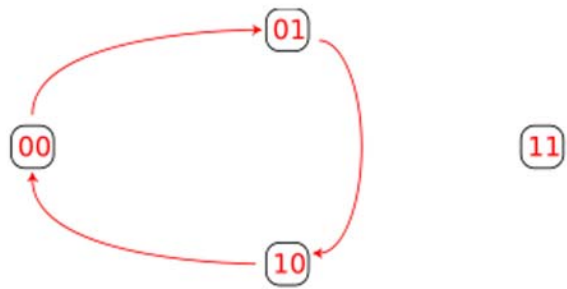

Figure 3: Binary de Bruijn graph for $m'=2$ showing only the cycle consisting of the last four states of $(1)$.

We demonstrate in detail that (and how) any sequence may be decomposed into a weighted superposition of such cycles, unique for each $m'$. Each cycle represents a different deterministic binary-state process, implicit in the fact that no state in such a cycle can be reached by, nor may it transition to, more than one other state. In the most general non-cyclic binary process, every $m'$-bit state may be reached from two preceding states and may transition to two. This follows from the fact that all transitions may be considered as the motion of a sliding window along the binary string: One digit is dropped at the beginning, one is added at the end. Given a state at time $t$, the state at time $t+1$ can end in only two possible states: the one ending in 0 or the one ending in 1. The converse is true as well reading the string in reverse order, hence any given state can be reached by two preceding ones. In ternary series, each state may transition to one of three possible states and may likewise be reached by three, and so on for higher bases.

Suppose a binary series represents transitions from one state to the next that are probabilistic. When a probabilistic process is decomposed into deterministic cycles, the transition probabilities between states are recast as transition probabilities among cycles at those states common to more than one cycle.

### 2.3 CYCLE DISTINCTIONS AMONG MG, MAJG AND $G

We here emphasize one of our important observations, namely, that just as the MG, MAJ and $G are distinct in terms of the persistence demonstrated by the time-series they generate; they are distinct as well in terms of their cycle structure.

**Figure 4** illustrates the typical weighting of cycles in the MG, MAJG and $G for m=4, $m'$=2, relative to the expected weighting in a completely random binary sequence. Bars in gray represent cycle weights in excess of the random expectation value normalized to 1; bars in black represent cycle weights less than 1.

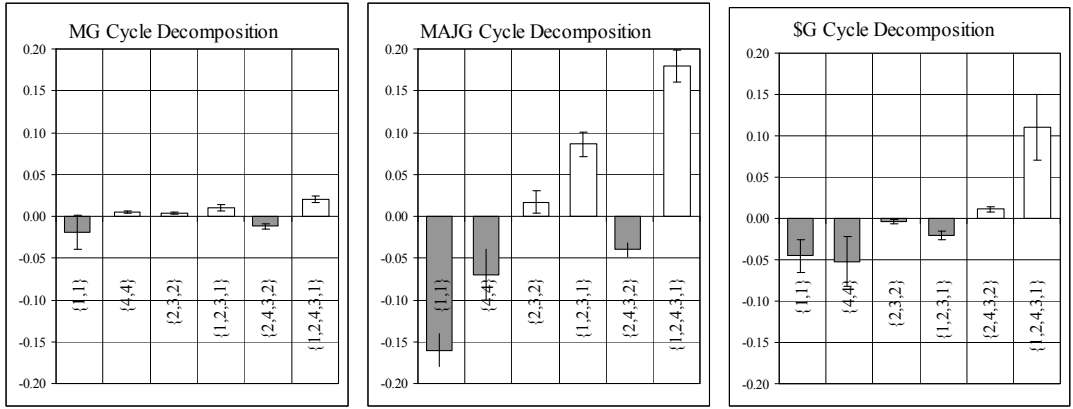

**Figure 4**: Relative weighting of cycles in the MG, MAJG and $G at m'=2 for *m*=*m^c*=4, *N*=31, *S*=2 expressed as a fraction/multiple of the expected weighting (normalized to 1) of a cycle decomposition of a completely random binary sequence

There are many potential way of both characterizing and applying such a "taxonomy" of series. One simple metric is the pseudo-Euclidian distance between the cycle decomposition of a given series and that for a completely random series, providing a measure of non-randomness. Taken exactly, such a measure presumes that cycles in a decomposition are orthonormal and they are not. But by normalizing the weights to 1 for a random series we get a somewhat better if still imprecise definition of "distance". Presuming orthogonality, this distance *d* is simply:

$$d_{game} = \sqrt{\sum_{j=1}^{k} w_j^2} \qquad (2)$$

As is evident from the absolute values of the distance from zero of each of the bars in **Figure 4**, $d_{maj} = 0.270 > d_S = 0.133 > d_{min} = 0.028$, i.e., the MAJG generates series that are, by this measure, the least random; the MG generates series that are closest to random. (A more exact use of this metric would require characterizing a series at many, theoretically all, $m'$.) The same relationship among the three games are illustrated in **Figure 1** with respect to persistence and anti-persistence: While the MG is anti-persistent and the MAJG and $Gs are persistent, the MG is closest to $\mathcal{P} = 0.5$, the random limit, the MAJG farthest.

Intuitively it might seem at first that anti-persistent sequences are less predictable than persistent sequences (from the perspective of an outside observer able to analyze a past history of arbitrary length). But in fact, the formal definition of persistence means that anti-persistent sequences of a given measure $\wp$ are as deterministic as persistent sequences of that same measure—so long as the $m'$ at which persistence is measured is consistent throughout. What **Figure 4** illustrates at a glance is that the overall departure from randomness of the MAJG and \$G is greater than for the MG and that the MAJG departs somewhat more from randomness than the \$G. On this basis we expect, and in fact find, that a prediction method based on a cycle decomposition should yield the best results for the MAJG and \$G relative to the MG.

### 2.4 CYCLE-BASED PREDICTORS FOR MG, MAJG AND \$G

We create a toy predictor (details are in Section 5.4) and apply it to all three games, pretending that the sequences they return are actual market price-changes. We find that for the MG at $m'=2$, the percent of correct predictions using a cycle-decomposition predictor is ~65%; for the MAJG the percent is ~72% and the \$G ~66%. Thus the predictability of the three game types by this test stand in the same relationship to one another as $d_{game}$ and $\wp_{game}$.

### 2.5 APPROXIMATION OF A SERIES AS A PERTURBED DOMINANT CYCLE

As discussed in ref. [3], the THMG, THMAJ and TH\$G may all be expressed analytically in the form of Markovian transition matrices. If these matrices are then decomposed into a weighted superposition of deterministic cycles, then the most heavily weighted cycle may be considered a zero[th] order approximation for the entire matrix. (In the limit of small $m$ and $\tau$, it may often happen that more than one cycle has the same, largest weight; these may be linked or disjoint.) As detailed in refs. [1,3] and below, the transition matrices for these games arises out of $N_U(t)$, the expression for the number of agents at timestep $t$ whose contribution to the collective state $A(t)$ needs to be determined by a coin toss (because their $S$ strategies have accumulated the same number of points but would yield at least two differing predictions), and $A_D(t)$, the expression for the contribution to the value of $A(t)$ that is wholly determined, contingent only upon the particular quenched disorder initializing the game and the particular "path history" at $t$, i.e., $\mu(t)$. A path-history is simply the union of the *m*-bit binary history at *t* (which we denote as $\mu_t$) and the preceding $\tau$-bit rolling window over which agent and strategy scores are maintained.

An agent's contribution to $A(t)$ will be determined when all of its strategies make the same prediction given $\mu(t)$. (This is more likely the smaller the value of *S*, i.e., the fewer the number of strategies per agent). This will happen only for some $\mu(t)$ if an agent's strategies differ, but of course for all $\mu(t)$ if they happen to be identical.

Now, $A(t) = A_D(t) + A_U(t)$, where $A_U(t)$ is the sum of all agent's contributions determined by an unbiased coin-toss. For certain $\mu(t)$ the absolute value of $A_D(t)$ may be large enough (a sufficient proportion of agents contributing to it) so that $A_U(t) < A_D(t)$—even if all the remaining agents (the number of which $= N_U(t)$) happen to vote the same way by chance and therefore generate the maximum possible $A_U(t)$. In these instances, $Sgn[A(t)] = Sgn[A_D(t)]$, i.e., the contribution to the series $\mu(t) \to \mu(t+1)$ will be determined and that step in the series is therefore deterministic. In terms of a transition matrix for a binary series, the entries representing the two transition probabilities from a given state to each of two possible successor states will consist of 1 and 0, implying that the transitions are either present or absent with certainty (i.e., wholly determined). In fact, as we will detail shortly, the dominant cycle in a decomposition of a Markovian transition matrix arises directly from the expression for $A_D(t)$, excluding the term $N_U(t)$. Its weight is therefore a direct measure of the degree of determinism present in the series.

A transition matrix presumes that a series is in fact Markovian. Series generated by the THMG, THMAJG and TH$G truly are. But the series generated by their non-TH variants, the MG, MAJG and $G proper are effectively Markovian only at equilibrium, and with intractably large matrices—the window of past information grows without bound and becomes equivalent to a sliding window only when very remote information no longer has an effect. However, as discussed in ref. [3], a very high-dimensional Markovian series may be approximated by a series of much lower dimension, capturing at least some of the information of the full matrix. Likewise may non-Markovian process be approximated, in a fashion similar to the use of hidden Markov models in which a more complex process is approximated by a hidden switching between two different Markovian processes [7][8]. (Indeed, the cycle decomposition method may be considered a simplification of the hidden Markov method in which the "switching" occurs among multiple deterministic matrices instead of between two probabilistic ones.)

The information contained in such an approximate transition matrix may be captured by analysis of a large window of preceding history. An efficient way of doing so is to create a cycle decomposition from this empirical data. If cycles emerge that are not merely somewhat but significantly weightier than all the others, one may hypothesize a significant degree of hidden determinism in the series as evidenced, for instance by the creation of predictors that successfully employ the weightiest element(s) of the decomposition.

Note, too, that in converting a continuously-valued (but time-discrete) series to a binary series, we are severely compressing the available data. Success or failure of a cycle-decomposition predictor in the teeth of real-world transaction costs provides a useful heuristic for the degree of data-preservation.

### 2.6 CYCLE-BASED PREDICTORS FOR THE NASDAQ COMPOSITE

Finally, we examine a major real-world financial series in these same terms, the NASDAQ composite (IXIC) over its entire history. We also examine the index divided in quarters to analyze the different performance characteristics of the predictor and relate it to varying degrees of persistence. We also incorporate a persistence filter with improved results.

We find that at $m'=2$, the overall degree of persistence of the IXIC is ~0.52, thus somewhat persistent like the series from the MAJG and $G and contrary to the MG. Its cycle decomposition (again over its entire history) is shown in **Figure 5**. (The cycle decomposition for each fourth of the series is presented later.) The cycle with weight

closest to zero is the last, Hamiltonian cycle $\{1,2,4,3,1\}$. The one-step persistent cycles $\{1,1\}$ and $\{4,4\}$ are over-represented relative to a random series and the one-step anti-persistent cycle $\{2,3,2\}$ is under-represented.

We thus have a snapshot that suggests a somewhat predictable because somewhat persistent time series. (The perfectly anti-persistent cycle, the last, is present with the lowest weight.) With $d_{IXIC} = 0.118$, the IXIC departs from randomness by this cycle measure to an extent midway between that of the MG and the $G.

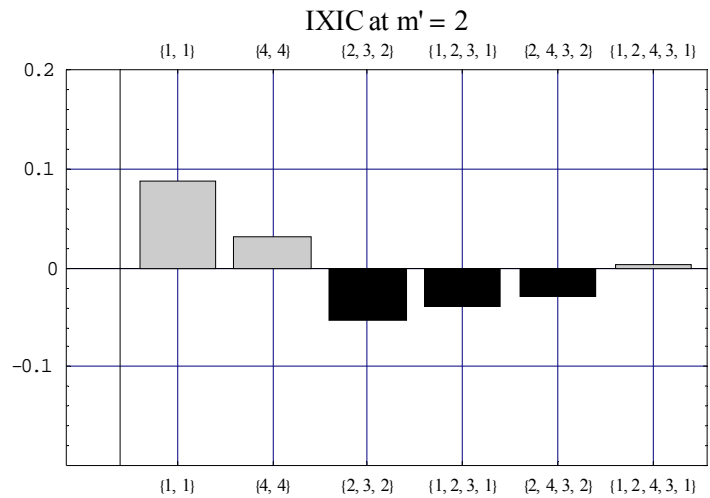

**Figure 5**: Cycle decomposition of the entire binary history of the NASDAQ composite index (IXIC) from February, 1971 through January, 2008.

From the over-representation in the IXIC of the two one-step persistent cycles one may guess informally at a simple strategy that has in fact been widely employed to trade the IXIC, namely the one-step Martingale: If the previous day's change in price is positive, buy (go long or hold); if negative, sell (or go short). Over the entire nearly 10,000 day history of the IXIC this strategy yields 58% correct directional guesses (which needs to be compared to a net upward drift over the history of 56% up days). This small degree of excess predictability, however, yields an annualized return of 38% per year (exclusive of transaction costs) versus a "buy and hold" return of 9% per year.. However, at 50 basis points per "round-trip" transaction (i.e., 50/10000 = .005 cost per unit value traded), the Martingale strategy would have yielded an annualized return of ~7%, more than 22% less than the buy and hold return of ~9%  A significant amount of information has nonetheless been detected by the Martingale strategy, but not enough to be of practical use—the predictor is highly inefficient, obtaining large gains at the cost of many changes in strategy direction. Clearly, insufficient information has been recovered from the binary-compressed price-change series to compensate for the cost of applying that information.

An explicit hidden (switching) Markov model of tick data used to generate a predictor for foreign exchange rates [9] yielded similar results: While significant theoretical predictability was obtained for the USD/CHF exchange rate, it was insufficient to overcome transaction costs, a common problem that plagues financial market predictors.

We do better by building a more precise (if still "toy") predictor based on the same cycle-decomposition for the IXIC index (NASDAQ composite; **Figure 5**), taking into account the entire cycle structure (in a simple way to be described) and again obtain a correct prediction percentage of 58% versus 56% up. The excess predictability in this case yields an annualized return of 0.36 versus the "buy and hold" return of 0.08—slightly worse than the Martingale predictor. However, returns remain superior to "buy and hold" for "round-trip" transaction costs of up to 50 basis points. Thus, the cycle predictor extracts information from the compressed binary time-series more efficiently than does the Martingale predictor—enough to warrant potential real world application.

Like most real-world series, the IXIC goes through periods of varying $\mathscr{P}$ (as measured over some time-scale). For example, we expect and find that during periods of low volatility as measured by proxy using the VIX (Chicago Board of Trade Volatility Index for the NASDAQ 100), the IXIC shows $\mathscr{P} > 0.5$, while during periods of high volatility, $\mathscr{P} > 0.5$. Furthermore, if recent past periods (50 trading days) of anti-persistence ($\mathscr{P} < 0.45$) are excluded (no market exposure on such days, either long or short), the returns (annualized over days of market exposure) rises to 0.38 (equivalent to the Martingale predictor) and remain superior to buy and hold for up to 60 basis points per round trip. In addition, since one is exposed to the market for only ~8% of trading days using this measure, monies are freed up for other uses the rest of the time.

Similar results are obtained for the US Dollar/Japanese Yen foreign exchange rate characterized by $\mathscr{P} \simeq 0.53$. However, if we apply these same methods to a real-world series with $\mathcal{P} \approx 0.5$, for example, the Philadelphia Exchange Gold and Silver Index (XAU), we find that they all fail.

It is worth comparing the cycle decomposition and its associated predictor to a simpler statistical analysis of dependencies and a comparable predictor we develop from these dependencies. In [10], Zhang studies the history of directional price changes in the NYSE Composite Index (NYA) of 400 stocks from 1966 to 1996. He displays the frequency that an up (+) daily price change occurs following each of the eight possible three-day sequences of directions of price changes. In **Figure 6** we duplicate his analysis of the NYA for the IXIC. (The figures are remarkably similar.)

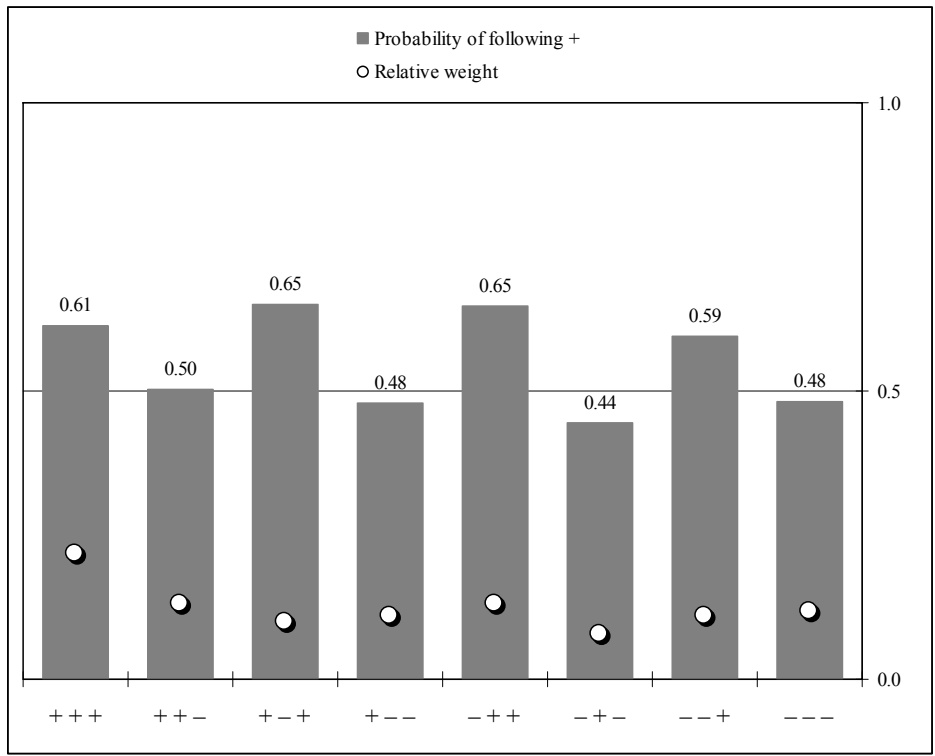

**Figure 6**: Frequency with which a positive daily change in price occurs following each of the eight possible three-day sequences of the direction of price-change for the IXIC. White circles indicate the relative weighting of the sequence. "+" = up; "−" = down. The small number of days of no change are excluded. (Daily data from February, 1971 through January, 2008)

In **Figure 7** we transform the results of **Figure 6** into expectations relative to a random sequence with the same overall upward bias as the NYA itself, to make it comparable to the cycle decomposition of **Figure 5**. **Figure 7** makes evident the rationale for the Martingale predictor: Regardless of the preceding two states, if the last state is "+" the next one is more likely to be as well; similarly with "−".

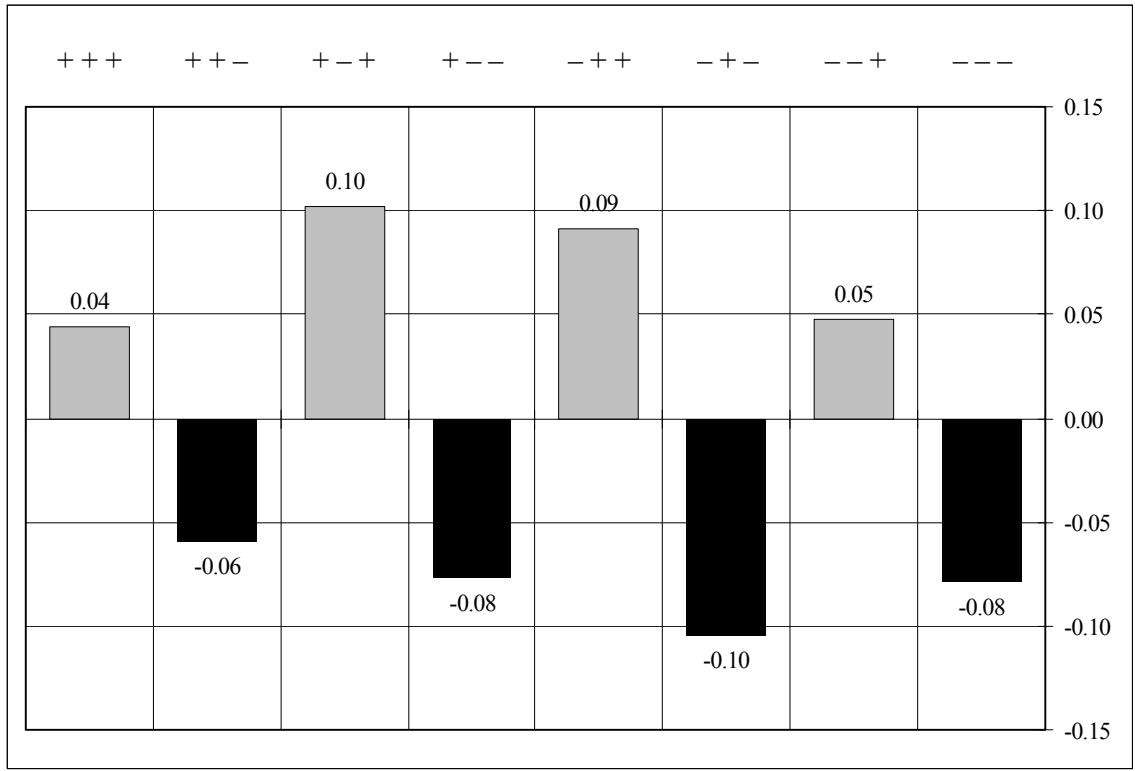

**Figure 7**: Relative probabilities of a positive daily price change following each of the 8 possible daily sequences of direction of price changes over three preceding days.

To improve the comparison—and highlight differences—the cycles of **Figure 5** may be recast according as follows. The cycle $\{4,4\}$ is equivalent to $11 \rightarrow 11 = 111 \equiv +++$, i.e., "$++$" followed by "$+$" with frequency (probability) of $0.50 + 0.03 = 0.53$. The $0.03$ comes from the second bar of **Figure 5**. The cycle $\{1,1\}$ is equivalent to $00 \rightarrow 00 = 000 \equiv ---$, i.e., "$--$" followed by "$-$" with frequency (probability) of $0.50 + 0.08 = 0.58$ (again from **Figure 5**) which is the same as followed by "$+$" with frequency (probability) of $1 - (0.50 + 0.08) = 0.42$. All the cycles of **Figure 5** may be recast according to **Table 1**. Note that not every possible sequence of states for any length other than 2 is included and that the selection of states based on the cycle decomposition includes states of four different lengths.

**Table 1**: Probability of + following state sequences equivalent to cycles

| Cycle | Prior States | Calculation | P(+) |
|---|---|---|---|
| {1,1} | $--$ | $1-(0.50+0.08)=$ | 0.42 |
| {4,4} | $++$ | $0.50+0.03=$ | 0.53 |
| {2,3,2} | $-+-$ | $0.50-0.06=$ | 0.44 |
| {1,2,3,1} | $--+-$ | $1-(0.50-0.04)=$ | 0.46 |
| {2,4,3,2} | $-++-$ | $0.50-0.03=$ | 0.47 |
| {1,2,4,3,1} | $--++-$ | $0.50+0.51=$ | 0.51 |

The results of **Table 1** are displayed in **Figure 8**.

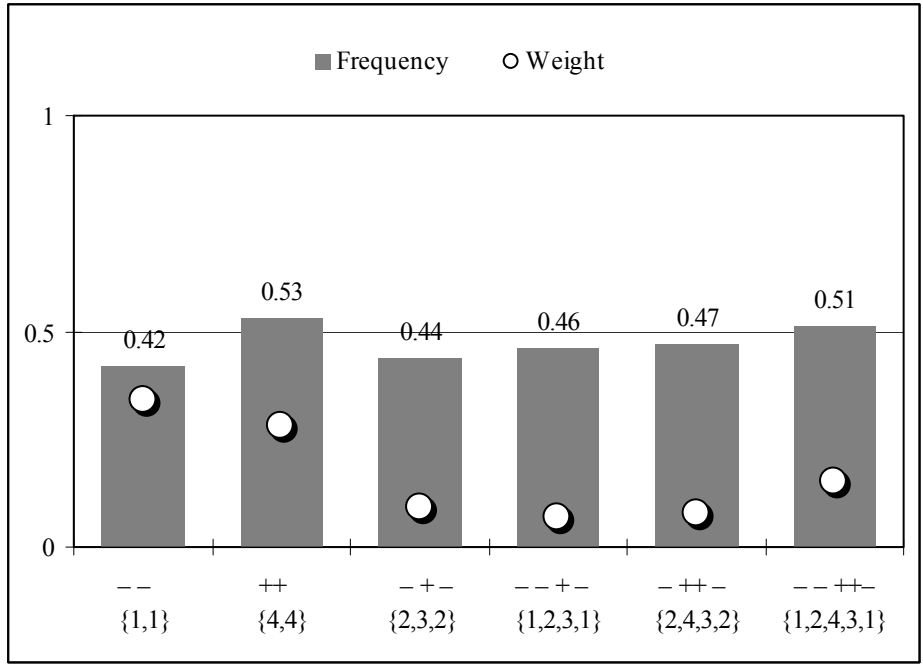

**Figure 8:** Cycle decomposition of the IXIC recast in terms of frequencies that a given sequence of daily directions of price changes is followed by a positive change in price.

The departures from 0.50 in **Figure 8** show mean dependencies over the entire history of the IXIC that are noticeably smaller than the departures and dependencies in **Figure 6**. At first glance therefore one might expect that a predictor based on the cycle decomposition would be less powerful than a comparable one based simply on the eight possible three-day histories. (The six sequences in the transformed cycle decomposition have a mean length of 3.3.) But the cycle decomposition arguably reveals more complex dependencies than the simpler statistical analysis.

In fact, using the eight three-day histories we construct a "dependency predictor" as close as possible in structure to the cycle-decomposition predictor and obtain the following results: Not taking into account transaction costs, the dependency predictor yields annualized returns of 26% — better than buy and hold (consistent with the arguments against the efficient market hypothesis made in [10]), but not so good as either the Martingale or cycle-decomposition predictor. (The superior performance of the Martingale predictor vis-à-vis the dependency predictor suggests that an approach based strictly on these dependencies may attain maximum performance at least for this application by utilizing histories of no longer than a single day.) Furthermore, while the cycle decomposition predictor retains performance superior to buy and hold (and to Martingale) for up to 50 basis points of transaction costs, dependency predictor gains—as well as effectively the entire initial investment—are wholly wiped out by such costs with an annualized loss of $-19\%$ per year. The dependency predictor retains performance superior to buy and hold only for transaction costs of no more than 17 basis points.

It is valuable to examine the performance of the predictor during a set of different arbitrarily selected time-periods. Following the procedure in [10] as applied to the NYA, we divide the IXIC price-change series in equal fourths and apply the cycle predictor to each. Results are shown in **Table 2** and in **Figure 9**.

**Table 2**: Summary of cycle predictor results for IXIC divided into equal fourths ("BH" = "Buy and Hold")

| Fourth | Dates | $\mathcal{P}$ | $<\mathcal{P}>$ | Ann.Ret.Cyc. | Ann.Ret.BH | Ann.Ret.Cyc−BH | $d$ |
|--------|-------|------|------|------|------|------|------|
| 1st | 2/5/71-5/2/80 | 0.56 | 0.55 | 57% | 2% | 55% | 0.16 |
| 2nd | 5/5/80-7/25/89 | 0.52 | 0.54 | 40% | 9% | 31% | 0.13 |
| 3rd | 7/26/89-10/14/98 | 0.52 | 0.52 | 49% | 23% | 26% | 0.11 |
| 4th | 10/15/98-1/23/08 | 0.49 | 0.49 | -7% | -6% | -1% | 0.09 |

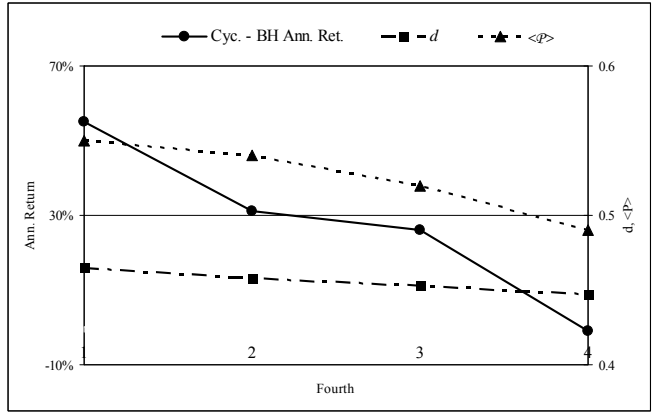

**Figure 9**: Annualized return of cycle predictor, by fourth of IXIC, in excess of buy and hold return showing correlation with $d$ and $<\mathcal{P}>$. $d$ is the pseudo-Euclidian distance between the cycle decomposition of the series and the expected decomposition of a random series with the same overall linear bias. $<\mathcal{P}>$ is the mean persistence of the series, i.e., averaged over $1 < m' < 10$.

For each fourth we determine the annualized raw return (exclusive of transaction costs) from the cycle predictor and the annualized buy and hold return. We compute the difference and denote the associated $\mathcal{P}$, $< \mathcal{P} >$ (averaged over $1 < m' < 10$) and $d$. We see that the raw return from the cycle predictor in excess of the buy and hold return declines monotonically over time with monotonic declines in $\mathcal{P}$, $< \mathcal{P} >$ and $d$. These results are consistent with the improved performance of the cycle predictor found previously with a persistence filter.

However, it is also striking that the capacity of the unmodified toy predictor to achieve gains in excess of buy and hold declines monotonically over the life of the IXIC. This suggests that as the NASDAQ market has matured, it has become more efficient. Improved results (in excess of buy and hold) are still found using the persistence filter suggesting not only that the increase in efficiency has not eliminated exploitable marginal inefficiencies but that the nature of these inefficiencies are related to the remaining degree (or periods) of persistence. This latter hypothesis is consistent with the overall monotonic decline in both $d$ and $<\mathcal{P}>$.

We now turn to a detailed explanation of the above findings.

# 3. MARKOV REPRESENTATION OF TH GAMES

## 3.1 DEFINITIONS AND PAYOFFS FOR THE MH-MG, -MAJG, AND -$G

In the simplest version of the MG, MAJG and $G, each of $N$ agents has $S = 2$ strategies, each of which encodes agents' action in response to a history of length $m = 2$. In the TH variants, the point (or score) table associated with strategies' success or failure is not maintained from the beginning of the game and is not ever growing. It is a rolling window of finite length $\tau$ (in the simplest case $\tau = 1$) ending at $t - m - 1$. At every discrete time-step $t$, each agent selects its action according to the following "optimization" rule: It acts in accord with that strategy which shows the greatest accumulation of "points"—the number of times the strategy would have won had it hypothetically chosen the same action as that selected by (1) the *minority* of all agents at that same time $t$ in the MG and THMG; or (2) the *majority* of all agents at that same time $t$ in the MAJG and THMAJG; or (3) the *majority* of all agents at time $t - 1$ in the $G and TH$G.

The differences among the three games whether in their standard or TH variant may be most easily summarized by the respective expressions for the gain/loss to agents or to strategies at each time step $t$, given the vote imbalance $A(t)$:

(1) For the MG/THMG: $\qquad g_i^{min}(t) = -a_i(t)A(t)$ or $g_i^{min}(t) = -Sgn\big[a_i(t)A(t)\big]$

(2) For the MAJG/THMAJG: $g_i^{maj}(t) = +a_i(t)A(t)$ or $g_i^{maj}(t) = -Sgn\big[a_i(t)A(t)\big]$

(3) For the $G/TH$G: $\qquad g_i^{\$}(t) = +a_i(t-1)A(t)$ or $g_i^{\$}(t) = +Sgn\big[a_i(t-1)A(t)\big]$

The subscript *i*, which indicates the agent number, may be replaced by "*i,s*" to indicate agent *i*'s $s^{th}$ strategy with $s \in \{1,2,...S\}$. We see from these expressions that in the MG, agents strive to avoid the majority (i.e., to garner a share of a scarce resource); in the MAJG they strive to join the majority (i.e., to follow a trend); and the $G they strive to have joined the majority in advance of the others (i.e., to anticipate turning points).

Because the standard MG is a frustrated system with multiple possible equilibria, it reaches an equilibrium state only after a very large but finite number of steps $t_{st}^{min}$. Both the MAJG and \$G attain their equilibrium states $t_{st}^{maj}$ and $t_{st}^{\$}$ rapidly. At $t_{st}$ the dynamics and the behavior of individual agents for a given initial quenched disorder state are indistinguishable from an otherwise identical game with $\tau \geq t_{st}$.

The fundamental result of the MG is generally cast in terms of system volatility: $\sigma^2/N$, that is, the fluctuations in the sign (or actual count) of the winning, minority state (i.e., the history). All variations of agent and strategy reward functions depend on the negative sign of the majority vote. Therefore both agent and strategy "wealth" (points, whether "real" or hypothetical) are inverse or negative (implicit) functions of the volatility: The lower the value of $\sigma^2/N$ the greater the mean "wealth" of the "system", i.e., of agents. To emphasize the relation of all three games to market-games and to optimization, we transform the fundamental results from statements on the properties of $\sigma^2/N$ to change in wealth, i.e., $\Delta w/\Delta t$ for agents and $\Delta w/\Delta t$ for strategies. These transformations of agent and strategy performance are intrinsically more natural for the MAJG and \$G.

We use the simplest possible computations— $g_i^{min}(t) = -Sgn\left[a_i(t)A(t)\right]$, $g_i^{maj}(t) = +Sgn\left[a_i(t)A(t)\right]$ and $g_i^{\$}(t) = +Sgn\left[a_i(t-1)A(t)\right]$—to track agent and strategy per-step gain.

### 3.2 DERIVATION OF THE FULL TRANSITION MATRIX

As discussed in [2] and previously in [1], the THMG—in contrast to the MG proper— may be expressed in Markov chain formalism. This formalism simplifies the analytic expression of the MG for a finite number of agents and lends itself for this reason more readily to real-world markets: Also because a finite-length history of past information corresponds more closely to the way that real-world trading strategies are devised (very old information is considered less pertinent than more recent information). Furthermore, there is no reason to believe that real-world markets are at or close to equilibrium. The Markov formulation and its decomposition into deterministic cycles is thereby naturally extended to real-world financial time-series.

The Markov formulation of the THMG requires (re-)partitioning the set agents into two groups at each time-step: A sub-set of agents whose decisions are wholly determined and the complementary sub-set of agents whose actions must be decided by chance. The relative dominance of the determined subset highlights the role of an underlying deterministic dynamic in the THMG to which the relatively rarer instances of random strategy selection may be viewed as a perturbation. Characterizing these two subsets analytically is therefore the means for developing a Markov formulation. This in turns allows decomposition of the Markov chain into a finite set of wholly deterministic cycles. The role of chance may then be characterized as jumps between cycles, each associated with a certain probability. The characteristic object of a Markov formulation is a transition matrix $\hat{\mathbf{T}}$ whose elements represent the probabilities of the system changing from one state to another. We here summarize only the elements of the Markov formalism for the THMG, THMAJG and TH\$G. Derivation and details may be found in [3].

At every discrete time-step $t$, each agent of $N$ (index $i$) independently re-selects one of its $S$ strategies. It "votes" as the selected strategy dictates by taking one of two "actions," designated by a binary value:

$$a_i(t) \in \{1, 0\}, \ \forall \ i, t \tag{2}$$

The state of the system as a whole at time $t$ is a mapping of the sum of all the agents' actions to the integer set $\{2N_1 - N\}$, where $N_1$ is the number of 1 votes and $N_0 = N - N_1$. This mapping is defined as :

$$A(t) = 2\sum_{i=1}^{N} a_i(t) - N = N_1 - N_0 \tag{3}$$

If $A(t) > \frac{N}{2}$, then the minority of agents will have chosen 0 at time $t$ ($N_0 < N_1$); if $A(t) < \frac{N}{2}$, then the minority of agents will have chosen 1 at time $t$ ($N_1 < N_0$). In the MG and THMG the minority choice is the "winning" decision for $t$. In the MAJG, THMAJG, \$G and TH\$G the majority choice is winning. This choice is then mapped back to $\{0, 1\}$ :

$$D_{sys}(t) = -\text{Sgn}\big[A(t)\big] \ \ \therefore D_{sys}(t) \in \{-1, +1\} \to \{0, 1\} \tag{4}$$

For the MG, MAJG and \$G, binary strings of length $m$ form histories $\mu(t)$, i.e., $m = \dim\left[\mu(t)\right]$. For the THMG, THMAJG and TH\$G binary strings of length $m + \tau$ form paths (or "path histories") [1,11], i.e., $m + \tau = \dim\left(\mu_t\right)$, where we define $\mu(t)$ as a history in the standard game and $\mu_t$ as a path in the TH game. Then as demonstrated in [1-3], any THMG has a Markov chain formulation. In our later discussion of the cycle decomposition method we will use the typical initial quenched disorder tensor $\hat{\Omega}$ shown in (5), and its symmetrized equivalent $\hat{\Psi} = \frac{1}{2}\left(\hat{\Omega} + \hat{\Omega}^{\top}\right)$ with $\{m, S, N\} = \{2, 2, 31\}$. Positions along all $S$ edges of $\hat{\Omega}$ represent an ordered listing of all available strategies. The numerical values $\Omega_{ij\ldots}$ in $\hat{\Omega}$ indicate the number of times a specific strategy-tuple has been selected in the initial endowment of the $S$ strategies to the $N$ agents. Without loss of generality, we may express $\hat{\Omega}$ in upper-triangular form since the order of strategies in a agent has no meaning. E.g., $\Omega_{2,5} = 3$ means that there are 3 agents comprised of strategy 2 and strategy 5.)

$$\hat{\Omega} = \begin{pmatrix} 1 & 2 & 0 & 0 & 1 & 1 & 0 & 0 \\ 0 & 0 & 0 & 0 & 3 & 3 & 1 & 1 \\ 0 & 0 & 2 & 0 & 1 & 0 & 0 & 0 \\ 0 & 0 & 0 & 1 & 1 & 0 & 0 & 1 \\ 0 & 0 & 0 & 0 & 1 & 0 & 2 & 1 \\ 0 & 0 & 0 & 0 & 0 & 2 & 2 & 1 \\ 0 & 0 & 0 & 0 & 0 & 0 & 2 & 1 \\ 0 & 0 & 0 & 0 & 0 & 0 & 0 & 0 \end{pmatrix} \tag{5}$$

Actions are drawn from a reduced strategy space (RSS) [12]. Each action is associated with a strategy $k$ and a path $\mu_t$. Together they can be represented in table form as a $\dim(\text{RSS}) \times \dim\left(\mu_t\right)$ binary matrix with elements converted for convenience from $\{0,1\} \rightarrow \{-1,+1\}$, i.e., $a_k^{\mu_t} \in \{-1,+1\}$ and defined as:

$$\hat{a} \equiv \begin{pmatrix} -1 & -1 & -1 & -1 \\ -1 & -1 & +1 & +1 \\ -1 & +1 & -1 & +1 \\ -1 & +1 & +1 & -1 \\ +1 & -1 & -1 & +1 \\ +1 & -1 & +1 & -1 \\ +1 & +1 & -1 & -1 \\ +1 & +1 & +1 & +1 \end{pmatrix} \tag{6}$$

The change in wealth (point gain or loss) associated with each of the 8 strategies for the 8 paths (= allowed transitions between the 4 histories) at any time $t$ is then:

$$\delta \bar{S}^{\min}_{\mu(t),\mu(t-1)} = +\left(\hat{a}^{\top}\right)_{\mu(t)} \times \left\{ 2Mod\left[\mu(t-1),2\right]-1 \right\} \tag{7}$$

$$\delta \bar{S}^{maj}_{\mu(t),\mu(t-1)} = -\left(\hat{a}^{\top}\right)_{\mu(t)} \times \left\{ 2Mod\left[\mu(t-1),2\right]-1 \right\} \tag{8}$$

$$\delta \bar{S}^{\$}_{\mu(t),\mu(t-1)} = -\left(\hat{a}^{\top}\right)_{\mu(t-1)} \times \left\{ 2Mod\left[\mu(t-1),2\right]-1 \right\} \tag{9}$$

($Mod\left[x,y\right]$ is "x modulo y"). $\mu(t)$ and $\mu(t-1)$ label each of the 4 histories $\{00,01,10,11\}$ hence take on one of values $\{1,2,3,4\}$. Equation (7) picks out from (6) the correct change in wealth over a single step since the strategies are ordered in symmetrical sequence.

The change in points associated with each strategy for each of the allowed transitions between paths $\mu_t$ of the last $\tau$ time steps used to score the strategies is:

$$\bar{s}^{game}_{\mu_t} \equiv \sum_{i=0}^{\tau-1} \delta \bar{S}^{game}_{\mu(t-i),\mu(t-i-1)} \tag{10}$$

For example, for $m = 2$ and $\tau = 1$ the strategy scores are kept for only a single time-step. There is no summation so (10) in matrix form reduces to:

$$\bar{s}^{game}_{\mu_t} \equiv \delta \bar{S}^{game}_{\mu(t),\mu(t-1)} \tag{11}$$

or, listing the results for all 8 path histories:

$$\hat{\mathbf{s}}_{\mu}^{game} \cong \delta\hat{\mathbf{S}}^{game} \tag{12}$$

$\delta\hat{\mathbf{S}}^{game}$ is an 8×8 matrix that can be read as a lookup table. It denotes the change in points accumulated over $\tau = 1$ time steps for each of the 8 strategies over each of the 8 path-histories

Instead of computing $A^{game}(t)$, we compute $A^{game}(\mu_t)$. Then for each of the $2^{m+\tau} = 8$ possible $\mu_t$, $A^{game}(\mu_t)$ is composed of a subset of wholly determined agent votes and a subset of undetermined agents whose votes must be determined by a coin toss:

$$A^{game}(\mu_t) = A_D^{game}(\mu_t) + A_U^{game}(\mu_t) \tag{13}$$

Some agents are undetermined at time t because their strategies have the same score and the tie has to be broken with an unbiased coin toss. The $A_U^{game}(\mu_t)$ are therefore random variables characterized by the binomial distribution. Their actual values vary with the number of undetermined agents, which may be expressed as [1-3]:

$$N_U^{min}(\mu_t) =$$
$$\left\{ \left( 1 - \left[ \left( \hat{a}^{\top} \right)_{\left( \text{Mod}[\mu_t - 1, 4] + 1 \right)} \otimes_{\delta} \left( \hat{a}^{\top} \right)_{\left( \text{Mod}[\mu_t - 1, 4] + 1 \right)} \right] \right) \circ \left( \bar{s}_{\mu_t}^{min} \otimes_{\delta} \bar{s}_{\mu_t}^{min} \right) \circ \hat{\mathbf{\Omega}} \right\}_{\left( \text{Mod}\left[ \mu_t - 1, 2^m \right] + 1 \right)} \tag{14}$$

$$N_U^{maj}(\mu_t) =$$
$$\left\{ \left( 1 - \left[ \left( \hat{a}^{\top} \right)_{\left( \text{Mod}[\mu_t - 1, 4] + 1 \right)} \otimes_{\delta} \left( \hat{a}^{\top} \right)_{\left( \text{Mod}[\mu_t - 1, 4] + 1 \right)} \right] \right) \circ \left( \bar{s}_{\mu_t}^{maj} \otimes_{\delta} \bar{s}_{\mu_t}^{maj} \right) \circ \hat{\mathbf{\Omega}} \right\}_{\left( \text{Mod}\left[ \mu_t - 1, 2^m \right] + 1 \right)} \tag{15}$$

$$N_U^{\$}(\mu_t) =$$
$$\left\{ \left( 1 - \left[ \left( \hat{a}^{\top} \right)_{\left( \text{Mod}[\mu_{t-1} - 1, 4] + 1 \right)} \otimes_{\delta} \left( \hat{a}^{\top} \right)_{\left( \text{Mod}[\mu_{t-1} - 1, 4] + 1 \right)} \right] \right) \circ \left( \bar{s}_{\mu_t}^{\$} \otimes_{\delta} \bar{s}_{\mu_t}^{\$} \right) \circ \hat{\mathbf{\Omega}} \right\}_{\left( \text{Mod}\left[ \mu_{t-1} - 1, 2^m \right] + 1 \right)} \tag{16}$$

"$\otimes_{\delta}$" is a generalized outer product, with the product being the Kronecker delta. $\bar{N}_U^{game}$ constitutes a vector of such values. The summed value of all undetermined decisions for a given $\mu_t$ is distributed binomially. Similarly [1-3]:

$$A_D^{min}(\mu_t) =$$
$$\left( \sum_{r=1}^{8} \left\{ \left[ \left( 1 - Sgn\left[ \bar{s}_{\mu_t}^{min} \ominus \bar{s}_{\mu_t}^{min} \right] \right) \circ \hat{\mathbf{\Psi}} \right] \bullet \hat{a} \right\}_r \right)_{\left( \text{Mod}\left[ \mu_t - 1, 2^m \right] + 1 \right)} \tag{17}$$

$$A_D^{maj}\left(\mu_t\right)=$$

$$\left(\sum_{r=1}^{8}\left\{\left[\left(1-Sgn\left[\,\bar{s}_{\mu_t}^{maj}\ominus\bar{s}_{\mu_t}^{maj}\right]\right)\circ\hat{\boldsymbol{\Psi}}\right]\bullet\left(-\hat{a}\right)\right\}_r\right)_{\left(\mathrm{Mod}\left[\mu_t-1,2^m\right]+1\right)}\tag{18}$$

$$A_D^{\$}\left(\mu_t\right)=$$

$$\left(\sum_{r=1}^{8}\left\{\left[\left(1-Sgn\left[\,\bar{s}_{\mu_t}^{\$}\ominus\bar{s}_{\mu_t}^{\$}\right]\right)\circ\hat{\boldsymbol{\Psi}}\right]\bullet\left(-\hat{a}\right)\right\}_r\right)_{\left(\mathrm{Mod}\left[\mu_{t-1}-1,2^m\right]+1\right)}\tag{19}$$

We define $\vec{A}_D^{game}$ as a vector of the determined contributions to $A(t)$ for each path $\mu_t$ in each respective game. In Eqs. (17)-(19), $\mu_t$ numbers paths from 1 to 8 and is therefore here an index. $\bar{s}_{\mu_t}^{game}$ is the "$\mu_t^{\text{th}}$" vector of net point gains or losses for each strategy when at $t$ the system has traversed the path $\mu_t$ ( i.e., it is the "$\mu_t^{\text{th}}$" element of the matrix $\hat{\mathbf{s}}_\mu^{game}=\delta\hat{\mathbf{S}}^{game}$ in (12)). "$\ominus$" is a generalized outer product of two vectors with subtraction as the product. The two vectors in this instance are the same, i.e., $\bar{s}_{\mu_t}^{game}$. "$\circ$" is Hadamard (element-by-element) multiplication and "$\bullet$" the standard inner product. The index $r$ refers to strategies in the RSS. Summation over $r$ transforms the base-ten code for $\mu_t$ into $\{1,2,3,4,1,2,3,4\}$. Selection of the proper number is indicated by the subscript expression on the entire right-hand side of (14). This expression yields an index number, i.e., selection takes place $1+$ Modulo 4 with respect to the value of $\left(\mu_t-1\right)$ in the THMG and THMAJG, and with respect to the value of $\left(\mu_{t-1}-1\right)$ in the TH\$G.

To obtain the transition matrix for the system as a whole, we require the $2^{m+\tau}\times 2^{m+\tau}$ adjacency matrix that filters out disallowed transitions. Its elements are $\Gamma_{\mu_t,\mu_{t-1}}$:

$$\hat{\boldsymbol{\Gamma}}=\begin{pmatrix}1 & 0 & 0 & 0 & 1 & 0 & 0 & 0\\1 & 0 & 0 & 0 & 1 & 0 & 0 & 0\\0 & 1 & 0 & 0 & 0 & 1 & 0 & 0\\0 & 1 & 0 & 0 & 0 & 1 & 0 & 0\\0 & 0 & 1 & 0 & 0 & 0 & 1 & 0\\0 & 0 & 1 & 0 & 0 & 0 & 1 & 0\\0 & 0 & 0 & 1 & 0 & 0 & 0 & 1\\0 & 0 & 0 & 1 & 0 & 0 & 0 & 1\end{pmatrix}\tag{20}$$

Equations (14)-(20) yield the history-dependent $(m+\tau) \times (m+\tau)$ matrix $\hat{\mathbf{T}}^{game}$ with elements $T_{\mu_t,\mu_{t-1}}^{game}$, representing the 16 allowed probabilities of transitions between the 2 sets of 8 path-histories $\mu_t$ and $\mu_{t-1}$:

$$T_{\mu_t,\mu_{t-1}}^{\min} =$$
$$\Gamma_{\mu_t,\mu_{t-1}} \times \qquad\qquad (21)$$
$$\sum_{x=0}^{N_U^{\min}(\mu_t)} \left\{ \mathbb{C}_x^{N_U^{\min}(\mu_t)} \left(\frac{1}{2}\right)^{N_U^{\min}(\mu_t)} \times \delta\left[ \mathrm{Sgn}\left(A_D^{\min}(\mu_t) + 2x - N_U^{\min}(\mu_t)\right) + \left(2\,\mathrm{Mod}\{\mu_{t-1},2\} - 1\right) \right] \right\}$$

$$T_{\mu_t,\mu_{t-1}}^{maj} =$$
$$\Gamma_{\mu_t,\mu_{t-1}} \times \qquad\qquad (22)$$
$$\sum_{x=0}^{N_U^{maj}(\mu_t)} \left\{ \mathbb{C}_x^{N_U^{maj}(\mu_t)} \left(\frac{1}{2}\right)^{N_U^{maj}(\mu_t)} \times \delta\left[ \mathrm{Sgn}\left(A_D^{maj}(\mu_t) + 2x - N_U^{maj}(\mu_t)\right) + \left(2\,\mathrm{Mod}\{\mu_{t-1},2\} - 1\right) \right] \right\}$$

$$T_{\mu_t,\mu_{t-1}}^{\$} =$$
$$\Gamma_{\mu_t,\mu_{t-1}} \times \qquad\qquad (23)$$
$$\sum_{x=0}^{N_U^{\$}(\mu_t)} \left\{ \mathbb{C}_x^{N_U^{\$}(\mu_t)} \left(\frac{1}{2}\right)^{N_U^{\$}(\mu_t)} \times \delta\left[ \mathrm{Sgn}\left(A_D^{\$}(\mu_t) + 2x - N_U^{\$}(\mu_t)\right) + \left(2\,\mathrm{Mod}\{\mu_{t-1},2\} - 1\right) \right] \right\}$$

The expression $\mathbb{C}_x^{N_{U,\mu_t}^{game}} \left(\frac{1}{2}\right)^{N_{U,\mu_t}^{game}}$ in (21)-(23) represents the binomial distribution of undetermined outcomes under a fair coin-toss required to break ties, with mean = $A_D^{game}(\mu_t)$.

As an example we will use later, for the quenched disorder matrix in (5), we obtain the following $\hat{\mathbf{T}}^{min}$:

$$\hat{\mathbf{T}}^{min} = \begin{pmatrix} 0 & 0 & 0 & 0 & 1 & 0 & 0 & 0 \\ 1 & 0 & 0 & 0 & 0 & 0 & 0 & 0 \\ 0 & 0 & 0 & 0 & 0 & 0 & 0 & 0 \\ 0 & 1 & 0 & 0 & 0 & 1 & 0 & 0 \\ 0 & 0 & 0.5 & 0 & 0 & 0 & 0.5 & 0 \\ 0 & 0 & 0.5 & 0 & 0 & 0 & 0.5 & 0 \\ 0 & 0 & 0 & 0.1445 & 0 & 0 & 0 & 1 \\ 0 & 0 & 0 & 0.8555 & 0 & 0 & 0 & 0 \end{pmatrix} \qquad (24)$$

## 3.3 EXTRACTION OF THE DOMINANT (DETERMINISTIC) CYCLE OR CYCLES

Suppose we were to round all entries in (24) to either 1 or 0. Such a matrix, assuming we could generate it un-ambivalently, should correspond to the most likely, wholly deterministic path traversed by the system. In the case (24) we cannot round off un-ambivalently as there are two pairs of entries equal exactly to 0.5. (This implies four possible deterministic configurations since choosing 1 or 0 for one entry in each pair automatically determines the value of the entry as pairs must add to 1):

$$\hat{\mathbf{T}}_0^{min} = \begin{pmatrix} 0 & 0 & 0 & 0 & 1 & 0 & 0 & 0 \\ 1 & 0 & 0 & 0 & 0 & 0 & 0 & 0 \\ 0 & 0 & 0 & 0 & 0 & 0 & 0 & 0 \\ 0 & 1 & 0 & 0 & 0 & 1 & 0 & 0 \\ 0 & 0 & ? & 0 & 0 & 0 & ? & 0 \\ 0 & 0 & 1-? & 0 & 0 & 0 & 1-? & 0 \\ 0 & 0 & 0 & 0 & 0 & 0 & 0 & 1 \\ 0 & 0 & 0 & 1 & 0 & 0 & 0 & 0 \end{pmatrix} \qquad (25)$$

(25) corresponds to the following definite transitions (read "column no. → row no."):

$$\begin{Bmatrix} 5 & \rightarrow & 1 \\ 1 & \rightarrow & 2 \\ 2 & \rightarrow & 4 \\ 6 & \rightarrow & 4 \\ 8 & \rightarrow & 7 \\ 4 & \rightarrow & 8 \end{Bmatrix} \qquad (26)$$

plus one of the following sets of additional transitions:

$$\begin{pmatrix} 3 & \to & 5 \\ 7 & \to & 5 \end{pmatrix}, \begin{pmatrix} 3 & \to & 5 \\ 7 & \to & 6 \end{pmatrix}, \begin{pmatrix} 3 & \to & 6 \\ 7 & \to & 5 \end{pmatrix}, \begin{pmatrix} 3 & \to & 6 \\ 7 & \to & 6 \end{pmatrix} \tag{27}$$

Note that there are no transitions *to* 3. 3 is an "emitting" state that will always be "absorbed" by other states should it appear (only once possibly, as the first state in the sequence). The two possible states which absorb it are 5 or 6, which deterministically go to 1 or 4 respectively. Thus, the only meaningful alternatives that the rounding procedure leaves us with are the transitions 7→5 or 7→6. These fill out (26) as follows:

$$\begin{pmatrix} 1 & \to & 2 \\ 2 & \to & 4 \\ 4 & \to & 8 \\ 8 & \to & 7 \\ 7 & \to & 5 \\ 5 & \to & 1 \end{pmatrix} or \begin{pmatrix} 7 & \to & 6 \\ 6 & \to & 4 \\ 4 & \to & 8 \\ 8 & \to & 7 \end{pmatrix} \tag{28}$$

corresponding to the cycles $\{1,2,4,8,7,5,1\}$ or $\{4,8,7,6,4\}$. From the second set of transitions in (28) we have excluded 1→2 and 2→4 as the sequence 1→2→4 is absorbed by $\{4,8,7,6,4\}$ in the same way that 3→5 or 3→6 are.

We anticipate from this intuitive analysis that there are two equiprobable deterministic cycles that can serve equally as the $0^{\text{th}}$ order approximation.

More formally, however, we can exclude from eqn. (21) the undetermined component, i.e., $N_U^{\min}(\mu_t)$ so that

$$T_{0,\mu_t,\mu_{t-1}}^{\min} =$$
$$\Gamma_{\mu_t,\mu_{t-1}} \times \delta\left[ \text{Sgn}\left(A_D^{\min}(\mu_t) + 2x\right) + \left(2\,\text{Mod}\{\mu_{t-1},2\} - 1\right) \right] \tag{29}$$

This equation yields the following transition matrix:

$$\hat{\mathbf{T}}_0^{\,min} = \begin{pmatrix} 0 & 0 & 0 & 0 & 1 & 0 & 0 & 0 \\ 1 & 0 & 0 & 0 & 0 & 0 & 0 & 0 \\ 0 & 0 & 0 & 0 & 0 & 0 & 0 & 0 \\ 0 & 1 & 0 & 0 & 0 & 1 & 0 & 0 \\ 0 & 0 & 0 & 0 & 0 & 0 & 0 & 0 \\ 0 & 0 & 0 & 0 & 0 & 0 & 0 & 0 \\ 0 & 0 & 0 & 0 & 0 & 0 & 0 & 1 \\ 0 & 0 & 0 & 1 & 0 & 0 & 0 & 0 \end{pmatrix} \tag{30}$$

It includes all deterministic transitions as in the rounding procedure but excludes the ones that are ambivalent (equiprobable). For quenched disorders that yield no entries of 0.5, the result is a single deterministic cycle. We will see that when the cycle decomposition is performed analytically on this $\hat{\mathbf{\Omega}}$ [1], the two equiprobable most heavily-weighted cycles reappear with the appropriate weight (and other less probable cycles appear with their appropriate weights). For most $\hat{\mathbf{\Omega}}$, especially those for larger $N$, equiprobable dominant cycles are unlikely.

## 4. CYCLE DECOMPOSITION OF TH-GAMES

In [4], the MG proper (not the THMG) is represented as a deterministic system with perturbations due to tie-breaks. The underlying deterministic system is likewise characterized as an Eulerian path on a de Bruijn graph even though the system's path is not Markovian. As described above, a close approximation of the underlying determinism in the THMG, THMAJG and TH\$G may be achieved simply by rounding every element of $\hat{\mathbf{T}}^{game}$ (eqns (21)-(23)) to either 0 or 1. When $N$ is relatively large, $\hat{\mathbf{T}}$ is relatively unlikely to have elements equal to $\frac{1}{2}$ and the rounded $\hat{\mathbf{T}}\,(\equiv \hat{\mathbf{T}}_0)$ well-characterizes the system, even for small $m$ and $\tau$. But for many interesting values of $N$ (i.e., $N = 31$), elements equal to $\frac{1}{2}$ are common in $\hat{\mathbf{T}}^{game}$ and there are different (binomially-distributed) $\hat{\mathbf{T}}_0^{\,game}$.

---

[1] So as not to obscure this potential complication, $\hat{\mathbf{\Omega}}$ was chosen in part because it illustrates it.

Note that elements rounded to 0 are eliminated from $\hat{\mathbf{T}}^{game}$. If no values other than 0,1 are allowed, $\hat{\mathbf{T}}_0^{game}$ will therefore either represent one cycle from the De Bruijn graph, or if more than one, the cycles will be disjoint. If elements exactly equal to $\frac{1}{2}$ are left unchanged, then for every such element there will be a cycle. The cycles will not necessarily be disjoint and the dynamic represented by $\hat{\mathbf{T}}_0^{game}$ may then be viewed as two (or more) deterministic orbits in state-space with a 0.5 probability of switching from one to another upon exiting any the state with element $\frac{1}{2}$. Similarly, $\hat{\mathbf{T}}^{game}$ unmodified (with arbitrary elements $T_{i,j}^{game} \in [0,1]$) can be thought of as a (probabilistically) weighted superposition of all possible deterministic cycles present in $\hat{\mathbf{T}}^{game}$. In general, the cycle with the greatest weight provides a good first-order approximation to $\hat{\mathbf{T}}^{game}$.

The upper limit on the number of terms (cycles) required to decompose a finite-sized transition matrix is also finite but grows super-exponentially with $m + \tau$. The accuracy of the approximation grows with the number of cycles included. But more important for our purposes is the fact that deterministic cycles are much easier to study individually than their composite, and characteristics emerge in the aggregate that shed light on features of the latter.

## 4.1 EXTRACTION OF CYCLES FROM A BINARY SERIES

We require first a method of extracting cycles from a given binary data series. Consider an arbitrary finite binary history (not necessarily derived from a TH $\hat{\mathbf{T}}^{game}$, but a purely arbitrary history selected at random):

$$\{1,0,0,1,0,0,0,1,1,1,0,1,1,1,0,1,0,0,1,1,0,0,0,1,1,0,1,1,0,0\ldots\} \tag{31}$$

This is equivalent to the following path-history (binary converted to digital) when paths are defined as having length $m' = 3$ (i.e., were eqn. (31) in fact a THMG history, then "$m + \tau$" would likewise = 3):

$$\{(1,0,0),(0,0,1),(0,1,0),(1,0,0),(0,0,0),(0,0,1),(0,1,1),(1,1,1),$$
$$(1,1,0),(1,0,1),(0,1,1),(1,1,1),(1,1,0),(1,0,1),(0,1,0),(1,0,0),$$
$$(0,0,1),(0,1,1),(1,1,0),(1,0,0),(0,0,0),(0,0,1),(0,1,1),(1,1,0),$$
$$(1,0,1),(0,1,1),(1,1,0),(1,0,0)\}\ldots \rightarrow \tag{32}$$
$$\{5,2,3,5,1,2,4,8,7,6,4,8,7,6,3,5,2,4,7,5,1,2,4,7,6,4,7,5\ldots\}$$

A cycle consists of any sequence of digital states of length $\leq 9 \left(= 2^{m'} + 1\right)$ that begins and ends with the same digit, and within which no digit is otherwise found more than once. The first cycle in this series is $\{5,2,3,5\}$. Extract and tabulate this cycle and replace it with 5, the repeated initial digit. The next cycle is $\{4,8,7,6,4\}$. Extract this cycle and replace it with 4. This leaves 5,1,2,4,8,7,6,3,5,2. Continue extracting and tabulating until no cycles are left. Ignore any remaining digits. Then repeat the process seven more times (i.e. for a total of $2^{m'}$ times), each time dropping one more digit from the beginning of the series. For the beginning of the series in (32) **Table 3** shows the first four of the eight required extractions.

**Table 3**: Extraction of Cycles from a Series

| Remaining Sequence | Extracted Cycles |
|---|---|
| (5,2,3,5,1,2,4,8,7,6,4,8,7,6,3,5,2,4,7,5,1,2,4,7,6,4,7,5) | – (5,2,3,5) |
| (5,1,2,4,8,7,6,4,8,7,6,3,5,2,4,7,5,1,2,4,7,6,4,7,5) | – (4,8,7,6,4) |
| (5,1,2,4,8,7,6,3,5,2,4,7,5,1,2,4,7,6,4,7,5) | – (5,1,2,4,8,7,6,3,5) |
| (5,2,4,7,5,1,2,4,7,6,4,7,5) | – (5,2,4,7,5) |
| (5,1,2,4,7,6,4,7,5) | – (4,7,6,4) |
| (5,1,2,4,7,5) | – (5,1,2,4,7,5) |
| (5) | |
| | |
| (2,3,5,1,2,4,8,7,6,4,8,7,6,3,5,2,4,7,5,1,2,4,7,6,4,7,5) | – (2,3,5,1,2) |
| (2,4,8,7,6,4,8,7,6,3,5,2,4,7,5,1,2,4,7,6,4,7,5) | – (4,8,7,6,4) |
| (2,4,8,7,6,3,5,2,4,7,5,1,2,4,7,6,4,7,5) | – (2,4,8,7,6,3,5,2) |
| (2,4,7,5,1,2,4,7,6,4,7,5) | – (2,4,7,5,1,2) |
| (2,4,7,6,4,7,5) | – (4,7,6,4) |
| (2,4,7,5) | |
| | |
| (3,5,1,2,4,8,7,6,4,8,7,6,3,5,2,4,7,5,1,2,4,7,6,4,7,5) | – (4,8,7,6,4) |
| (3,5,1,2,4,8,7,6,3,5,2,4,7,5,1,2,4,7,6,4,7,5) | – (3,5,1,2,4,8,7,6,3) |
| (3,5,2,4,7,5,1,2,4,7,6,4,7,5) | – (5,2,4,7,5) |
| (3,5,1,2,4,7,6,4,7,5) | – (4,7,6,4) |
| (3,5,1,2,4,7,5) | – (5,1,2,4,7,5) |
| (3,5) | |
| (5,1,2,4,8,7,6,4,8,7,6,3,5,2,4,7,5,1,2,4,7,6,4,7,5) | – (4,8,7,6,4) |
| (5,1,2,4,8,7,6,3,5,2,4,7,5,1,2,4,7,6,4,7,5) | – (2,4,8,7,6,3,5,2) |
| (5,1,2,4,7,5,1,2,4,7,6,4,7,5) | – (1,2,4,7,5,1) |
| (5,1,2,4,7,6,4,7,5) | – (4,7,6,4) |
| (5,1,2,4,7,5) | – (5,1,2,4,7,5) |
| (5) | |

Categorize and count the number of times each cycle appears and compute its proportion as a fraction all cycle types. Adjust this proportion by treating cyclic permutations of a cycle as the same type and combine them into one category. Results for the partial series (four of eight extractions) in (32) are shown in **Table 4**. The proportional representation of a cycle in the *complete* extraction constitutes its "weight".

**Table 4**: Computation of Cycle Weights

| Cycle | Raw Count | Adj. Count | Weight (Proportion) |
|---|---|---|---|
| (4,7,6,4) | 4 | 4 | 0.190 |
| (5,2,3,5) | 1 | 1 | 0.048 |
| (2,3,5,1,2) | 1 | 1 | 0.048 |
| (4,8,7,6,4) | 4 | 4 | 0.190 |
| (5,2,4,7,5) | 2 | 2 | 0.095 |
| (1,2,4,7,5,1) | 1 | 5 | 0.238 |
| (2,4,7,5,1,2) | 1 | – | – |
| (5,1,2,4,7,5) | 3 | – | – |
| (2,4,8,7,6,3,5,2) | 2 | 2 | 0.095 |
| (3,5,1,2,4,8,7,6,3) | 1 | 2 | 0.095 |
| (5,1,2,4,8,7,6,3,5) | 1 | – | – |

The cycles in column 1 of **Table 4** can be represented as adjacency matrices within the space of all possible cycles on the binary deBruijn graph of order 3 as shown in **Figure 10**. Call this column vector $\bar{\mathbf{J}}$.

**Figure 10**: $\bar{\mathbf{J}}$ = the 8 non-zero-weight cycles of (32) in adjacency matrix form.

Call column 4 of **Table 4** $\bar{\omega}$, the weights of every cycle. The dot product $\bar{\mathbf{J}} \cdot \bar{\omega}$ yields (33), the transition matrix for the series in (32):

$$\bar{\mathbf{J}} \cdot \bar{\omega} = \begin{pmatrix} 0 & 0 & 0 & 0 & 0.38 & 0 & 0 & 0 \\ 0.38 & 0 & 0 & 0 & 0.24 & 0 & 0 & 0 \\ 0 & 0.10 & 0 & 0 & 0 & 0.19 & 0 & 0 \\ 0 & 0.52 & 0 & 0 & 0 & 0.38 & 0 & 0 \\ 0 & 0 & 0.29 & 0 & 0 & 0 & 0.33 & 0 \\ 0 & 0 & 0 & 0 & 0 & 0 & 0.57 & 0 \\ 0 & 0 & 0 & 0.52 & 0 & 0 & 0 & 0.38 \\ 0 & 0 & 0 & 0.38 & 0 & 0 & 0 & 0 \end{pmatrix} \qquad (33)$$

In this case the resulting matrix has complementary entries that do not add to 1 because the series is so short. But as proven formally in [13,14], in the limit of a sufficiently long such series, the transition values obtained by a complete extraction are accurate and arbitrarily precise, i.e., for every pair of non-zero entries $\left\{ J_{k_{i,j}}, J_{k_{i+1,j}} \right\}$ representing an allowed transition in the $R \times R$ transition matrix $J_k$ (where $\left\{ J_1, J_2, \ldots J_k, \ldots J_R \right\} \equiv \bar{\mathbf{J}}$), $\lim_{L \to \infty} \left( J_{k_{i,j}} + J_{k_{i+1,j}} \right) = 1$, with $L$ the series' length. In any series of such finite length, the complementary entries $\left\{ J_{k_{i,j}}, J_{k_{i+1,j}} \right\}$ may be normalized so their sum equals 1 to convert actual transition frequencies to approximate transition probabilities:

$$\begin{pmatrix} 0 & 0 & 0 & 0 & 0.38 & 0 & 0 & 0 \\ 0.38 & 0 & 0 & 0 & 0.24 & 0 & 0 & 0 \\ 0 & 0.10 & 0 & 0 & 0 & 0.19 & 0 & 0 \\ 0 & 0.52 & 0 & 0 & 0 & 0.38 & 0 & 0 \\ 0 & 0 & 0.29 & 0 & 0 & 0 & 0.33 & 0 \\ 0 & 0 & 0 & 0 & 0 & 0 & 0.57 & 0 \\ 0 & 0 & 0 & 0.52 & 0 & 0 & 0 & 0.38 \\ 0 & 0 & 0 & 0.38 & 0 & 0 & 0 & 0 \end{pmatrix} \rightarrow \begin{pmatrix} 0 & 0 & 0 & 0 & 0.61 & 0 & 0 & 0 \\ 1 & 0 & 0 & 0 & 0.39 & 0 & 0 & 0 \\ 0 & 0.16 & 0 & 0 & 0 & 0.33 & 0 & 0 \\ 0 & 0.84 & 0 & 0 & 0 & 0.67 & 0 & 0 \\ 0 & 0 & 1 & 0 & 0 & 0 & 0.37 & 0 \\ 0 & 0 & 0 & 0 & 0 & 0 & 0.63 & 0 \\ 0 & 0 & 0 & 0.58 & 0 & 0 & 0 & 1 \\ 0 & 0 & 0 & 0.42 & 0 & 0 & 0 & 0 \end{pmatrix} \qquad (34)$$

Note, too, that the weightiest cycle treated as having weight = 1 provides a $0^{\text{th}}$ order (and therefore wholly deterministic) approximation for the series. (The closer its actual weight is to 1 the more accurate the approximation) . In the above instance:

$$\begin{pmatrix} 0 & 0 & 0 & 0 & 1 & 0 & 0 & 0 \\ 1 & 0 & 0 & 0 & 0 & 0 & 0 & 0 \\ 0 & 0 & 0 & 0 & 0 & 0 & 0 & 0 \\ 0 & 1 & 0 & 0 & 0 & 0 & 0 & 0 \\ 0 & 0 & 0 & 0 & 0 & 0 & 1 & 0 \\ 0 & 0 & 0 & 0 & 0 & 0 & 0 & 0 \\ 0 & 0 & 0 & 1 & 0 & 0 & 0 & 0 \\ 0 & 0 & 0 & 0 & 0 & 0 & 0 & 0 \end{pmatrix} \quad (35)$$

We may create higher order (no longer wholly deterministic) approximations by adding the second weightiest matrix, third weightiest, etc. (and again normalizing the entries to 1).

We now demonstrate the use of this decomposition in the series generated by the THMG by analytically decomposing the transition matrix $\hat{\mathbf{T}}^{min}$ derived from an initial quenched disorder tensor $\hat{\Omega}$, in particular the tensor of (5). Identical methods will decompose $\hat{\mathbf{T}}^{maj}$ and $\hat{\mathbf{T}}^{\$}$. We will later use the above empirical method to approximate series generated by the (non-TH) MG, MAJG and $G as well as real-world series. Note that the decomposition of a given finite series produces an approximation of a transition matrix whereas an actual transition matrix generates a series of unlimited length.

## 4.2 ANALYTIC FORM OF A CYCLE DECOMPOSITION

The decomposition of a known transition matrix into its weighted cycle structure was first proposed in Ref. [13]. For any stochastic matrix $\hat{\mathbf{T}}$ (of which the transition matrix for THMG is an instance of size $2^{m+\tau} \times 2^{m+\tau}$), and the set $C$ of all possible directed cycles $J^c$ of the adjacency matrix $\hat{\Gamma}$ of $\hat{\mathbf{T}}$, it can be shown that [15]:

$$T_{i,j} = \frac{\sum_c \omega_c J_{i,j}^c}{\sum_c \omega_c J_i^c}, \quad 1 \le c \le \dim(C), \quad 1 \le i, j \le 2^{m+\tau} \quad (36)$$

(recall that the $\omega_c$ are the cycle weights) where

$$J_i^c = \sum_{j=1}^{2^{m+\tau}} J_{i,j}^c = \sum_{i=1}^{2^{m+\tau}} J_{i,j}^c. \quad (37)$$

$J_{i,j}^c = 1$ if $(i,j)$ is a directed edge of $J^c$, 0 otherwise. In addition,

$$\pi_i = \sum_c \omega_c J_i^c \qquad (38)$$

so that

$$\pi_i T_{i,j} = \sum_c \omega_c J_{i,j}^c; \quad \hat{\mathbf{T}} \cdot \bar{\pi} = \bar{\pi} = \sum_c J^c \cdot \bar{\omega} \qquad (39)$$

where the $\pi_i$ are the $2^{m+\tau}$ steady-state probabilities derivable from $\hat{\mathbf{T}}$. (39) represents dim($C$) simultaneous matrix equations to be solved for the $\omega_c$.

For the $\hat{\mathbf{T}}^{min}$ of equation (5) ($\hat{\Gamma}$ of equation (20)) which has 4 cycles only, and which has $S$ agents only,

$$\bar{\mathbf{J}} \equiv \left\{ J^1, J^2, J^3, J^4 \right\} = \qquad (40)$$

$$\left\{
\begin{pmatrix}
0 & 0 & 0 & 0 & 0 & 0 & 0 & 0 \\
0 & 0 & 0 & 0 & 0 & 0 & 0 & 0 \\
0 & 0 & 0 & 0 & 0 & 0 & 0 & 0 \\
0 & 0 & 0 & 0 & 0 & 1 & 0 \\
0 & 0 & 0 & 0 & 0 & 0 & 0 & 0 \\
0 & 0 & 1 & 0 & 0 & 0 & 0 & 0 \\
0 & 0 & 0 & 0 & 0 & 1 & 0 & 0 \\
0 & 0 & 0 & 0 & 0 & 0 & 0 & 0
\end{pmatrix},
\begin{pmatrix}
0 & 0 & 0 & 0 & 0 & 0 & 0 & 0 \\
0 & 0 & 0 & 0 & 0 & 0 & 0 & 0 \\
0 & 0 & 0 & 0 & 0 & 0 & 0 & 0 \\
0 & 0 & 0 & 0 & 0 & 0 & 0 & 1 \\
0 & 0 & 0 & 0 & 0 & 0 & 0 & 0 \\
0 & 0 & 0 & 1 & 0 & 0 & 0 & 0 \\
0 & 0 & 0 & 0 & 0 & 1 & 0 & 0 \\
0 & 0 & 0 & 0 & 0 & 0 & 1 & 0
\end{pmatrix},
\begin{pmatrix}
0 & 1 & 0 & 0 & 0 & 0 & 0 & 0 \\
0 & 0 & 0 & 1 & 0 & 0 & 0 & 0 \\
0 & 0 & 0 & 0 & 0 & 0 & 0 & 0 \\
0 & 0 & 0 & 0 & 0 & 1 & 0 \\
1 & 0 & 0 & 0 & 0 & 0 & 0 & 0 \\
0 & 0 & 0 & 0 & 0 & 0 & 0 & 0 \\
0 & 0 & 0 & 1 & 0 & 0 & 0 & 0 \\
0 & 0 & 0 & 0 & 0 & 0 & 0 & 0
\end{pmatrix},
\begin{pmatrix}
0 & 1 & 0 & 0 & 0 & 0 & 0 & 0 \\
0 & 0 & 0 & 1 & 0 & 0 & 0 & 0 \\
0 & 0 & 0 & 0 & 0 & 0 & 0 & 0 \\
0 & 0 & 0 & 0 & 0 & 0 & 0 & 1 \\
1 & 0 & 0 & 0 & 0 & 0 & 0 & 0 \\
0 & 0 & 0 & 0 & 0 & 0 & 0 & 0 \\
0 & 0 & 0 & 1 & 0 & 0 & 0 \\
0 & 0 & 0 & 0 & 0 & 0 & 1 & 0
\end{pmatrix}
\right\}$$

corresponding to $\left\{ (4,7,6,4), (4,8,7,6,4), (1,2,4,7,5,1), (1,2,4,8,7,5,1) \right\}$ and their cyclic permutations. (In general, a random binary series converted to paths of length 3 can and if long enough will have 19 unique cycles. The truncated series (32) we used as a sample contains only the 8 cycles shown in adjacency matrix form in **Figure 10**, of which we showed the extraction of 4 in detail. In the present example derived from a THMG tensor, there are only 4 extractable cycles in toto reflecting the high degree of determinism in the time series generated by $\hat{\Omega}$. From (37):

$$\hat{\mathbf{J}} \equiv \begin{pmatrix} J_1^1 & \cdots & J_1^4 \\ \vdots & \ddots & \vdots \\ J_8^1 & \cdots & J_8^4 \end{pmatrix} = \begin{pmatrix} 0 & 0 & 1 & 1 \\ 0 & 0 & 1 & 1 \\ 0 & 0 & 0 & 0 \\ 1 & 1 & 1 & 1 \\ 0 & 0 & 1 & 1 \\ 1 & 1 & 0 & 0 \\ 1 & 1 & 1 & 1 \\ 0 & 1 & 0 & 1 \end{pmatrix} \tag{41}$$

Solving (39) (equivalently, $\bar{\pi} = \hat{\mathbf{J}} \cdot \bar{\omega}$ ), for the $\omega_c$ using the method of cofactor expansion, we obtain the approximate values shown in **Table 5**.

<div align="center">

**Table 5**: Cycle Weights

|  | Cycle | *Weight* |
|---|---|---|
| $\omega_1$ | (4,7,6,4) | 0.072 |
| $\omega_2$ | (4,8,7,6,4) | 0.428 |
| $\omega_3$ | (1,2,4,7,5,1) | 0.072 |
| $\omega_4$ | (1,2,4,8,7,5,1) | 0.428 |

</div>

To check on the correctness of the solution for equation (20), $\vec{\mathbf{J}} \cdot \bar{\omega} \equiv \hat{\mathbf{T}}_{cyc}$ :

$$\hat{\mathbf{T}} = \begin{pmatrix} 0 & 0 & 0 & 0 & 1 & 0 & 0 & 0 \\ 1 & 0 & 0 & 0 & 0 & 0 & 0 & 0 \\ 0 & 0 & 0 & 0 & 0 & 0 & 0 & 0 \\ 0 & 1 & 0 & 0 & 0 & 1 & 0 & 0 \\ 0 & 0 & 0.5 & 0 & 0 & 0 & 0.5 & 0 \\ 0 & 0 & 0.5 & 0 & 0 & 0 & 0.5 & 0 \\ 0 & 0 & 0 & 0.1445 & 0 & 0 & 0 & 1 \\ 0 & 0 & 0 & 0.8555 & 0 & 0 & 0 & 0 \end{pmatrix}; \hat{\mathbf{T}}_{cyc} = \begin{pmatrix} 0 & 0 & 0 & 0 & 1 & 0 & 0 & 0 \\ 1 & 0 & 0 & 0 & 0 & 0 & 0 & 0 \\ 0 & 0 & 0 & 0 & 0 & 0 & 0 & 0 \\ 0 & 1 & 0 & 0 & 0 & 1 & 0 & 0 \\ 0 & 0 & 0 & 0 & 0 & 0 & 0.5 & 0 \\ 0 & 0 & 0 & 0 & 0 & 0 & 0.5 & 0 \\ 0 & 0 & 0 & 0.1445 & 0 & 0 & 0 & 1 \\ 0 & 0 & 0 & 0.8555 & 0 & 0 & 0 & 0 \end{pmatrix} \tag{42}$$

Note the one pair of entries that differ between the two matrices in (42). The matrix equation (39) may be solved by a number of methods (e.g., division-free row reduction, one-step row reduction; we used cofactor expansion). All typically also have difficulty with probabilities = 0.5. We see that this has happened here. Formally, the convergence of $\hat{\mathbf{T}}_{cyc}$ to $\hat{\mathbf{T}}$ is in general weak and "almost sure" [16]. We will see by comparison to numerical simulation that the small error thus introduced has no effect on our use of the decomposition. The difference between the two matrices in (42) provides an illustration of one reason why this has no effect: The original matrix $\hat{\mathbf{T}}$ on the left of (42) shows an equiprobable transition from path 3 [from$(0,1,0)$] to path 5 or to path 6 [to $(1,0,0)$ or to $(1,0,1)$]. But looking at row 3 of either matrix, we see that state 3 is inaccessible altogether (no transitions from any other state to it), hence no transition *from* state 3 is possible. This is reflected in $\hat{\mathbf{T}}_{cyc}$ (on the right) by the two 0 zero entries in place of 0.5. Thus, the two matrices, being at equilibrium, are equivalent.

### 4.3 COMPARISON TO NUMERICAL SIMULATION

Each cycle in the decomposition is a purely deterministic, discrete periodic process. As noted before, the elements of $\hat{\mathbf{T}}_{cyc}$ represent the probabilities of transitions between cycles rather than (equivalently in $\hat{\mathbf{T}}$) transitions between path histories. All the dynamics of the full THMG can therefore be derived from the deterministic dynamics of the THMG around just one of every cycle, properly composed and weighted. For example, if $\hat{\mathbf{T}}_{cyc}$ were composed of just two cycles of respective lengths 5 and 4, the total change in agent and strategy wealth over these will be computed over twenty (the lowest common multiple) steps—4 of the first cycle times 5 of the second, each times its appropriate weight, and then summed. The net per-step gain (or loss) in wealth for both agents and for their underlying strategies is easily computable around any cycle using (7) -(12). For the $\hat{\mathbf{T}}$ and $\hat{\mathbf{T}}_{cyc}$ of (20), with 31 agents, overall results are shown in **Table 6**.

**Table 6**: Elements of the Cycle Decomposition of the THMG with S and C agents

| Cycle | Weight | ΔW |
|---|---|---|
| **(4,7,6,4)** | | |
| $\Delta W_{agent}$ | *0.08(4)* | −0.10 |
| $\Delta W_{strategy}$ | | −0.09 |
| **(4,8,7,6,4)** | | |
| $\Delta W_{agent}$ | *0.85(4)* | −0.23 |
| $\Delta W_{strategy}$ | | −0.06 |
| **(1,2,4,7,5,1)** | | |
| $\Delta W_{agent}$ | *0.06(3)* | −0.11 |
| $\Delta W_{strategy}$ | | −0.04 |
| **(1,2,4,8,7,5,1)** | | |
| $\Delta W_{agent}$ | *0.00(0)* | −0.19 |
| $\Delta W_{strategy}$ | | −0.03 |
| ***Weighted Composite*** | | |
| $\Delta W_{agent}$ | ***1.00(0)*** | **−0.21** |
| $\Delta W_{strategy}$ | | **−0.06** |

Comparison to numerical and the standard analytic results are in **Table 7** showing the close agreement among methods. ("Analytic" results here refer to the mean per-step change in wealth averaged over all agents and all strategies respectively, i.e., $\Delta W_{agent}, \Delta W_{strategy}$ as detailed in section 3.2 as well as in [2] and [1]. Numeric results refer to numeric simulations of the THMG. In both cases we refer here to a single initial quenched disorder matrix. As explained in [2], this example is typical with changes in wealth close to the average over many initial quenched disorders.)

**Table 7**: Comparison of cycle decomposition to analytic and numeric results

|  | $\Delta W_{agent}$ | $\Delta W_{strategy}$ |
|---|---|---|
| *Numeric* | −0.20 | −0.07 |
| *Analytic* | −0.22 | −0.06 |
| ***Cycle Decomp.*** | −0.21 | −0.06 |

## 4.4 CYCLE ANALYSIS OF THE ILLUSION OF CONTROL

Ref. [2] discusses another feature of MG's and THMG's: Even though agents' choice of strategy is designed to optimize their performance, the performance of agents (their mean per-step accumulation of wealth) is always on average poorer than the measured performance of their $S$ strategies. Agents choose which strategy to use based on a comparison of how their strategies perform: The optimization rule employed is simply "deploy whichever strategy would have accumulated the largest number of points over the time horizon $\tau$". Yet by this measure agents appear to do better by selecting a single strategy and sticking to it, or by randomly selecting strategies (or actions) at each time-step. Thus the control method in the MG and THMG yields what may be called an "illusion of control". The relative outperformance of strategies as thus defined can be seen as arising from the anti-persistence characteristic of MG-generated time-series. As discussed in ref. [3], MAJG, THMAJG, $G and TH$G display no illusion of control and the time-series they generate are persistent at all $m$.

Another way of understanding this phenomenon is by examining the performance of agents and strategies around deterministic cycles in the THMG. Depending on the weights, a single cycle may dominate the behavior of the THMG. In other instances just a few cycles dominate. It turns out that, for a majority of $\hat{\Omega}$, most cycles show an illusion of control under the MG optimization rule for most strategy pairs. Therefore, the illusion is expected on average over many different initial configurations of $\hat{\Omega}$. As cycles are deterministic, we can examine the behavior of the optimization rule in particularly simple form. **Table 8** illustrates the behavior of the MG optimization rule around all possible (allowed) cycles as averaged over every possible strategy and every agent (i.e., every possible pair of strategies) for $\{m, S, N, \tau\} = \{2, 2, 31, 1\}$. (This is the first occasion where [almost] all possible cycles based on paths of length 3 are listed. We take this opportunity to note that the two simplest cycles, $(1,1)$ and $(8,8)$, and only these two, are absent. This is a typical feature of the THMG and demonstrates immediately the tendency of its binary time-series toward anti-persistence, at least at the smallest possible $m'$.)

**Table 8**: Optimization of All Standard Strategy Pairs over All Cycles and time steps

| Cycle | $\langle \Delta W_{agent,\,all\,pairs} \rangle$ per cycle | $\langle \Delta W_{agent,\,all\,pairs} \rangle$ per time-step | $\langle \Delta W_{strats,\,all} \rangle$ per cycle | $\langle \Delta W_{strats,\,all} \rangle$ per time-step |
|---|---|---|---|---|
| (3,6,3) | 0. | 0. | 0. | 0. |
| (2,3,5,2) | 0. | 0. | 0. | 0. |
| (4,7,6,4) | 0. | 0. | 0. | 0. |
| (1,2,3,5,1) | –0.5 | –0.125 | 0. | 0. |
| (2,4,7,5,2) | 0. | 0. | 0. | 0. |
| (4,8,7,6,4) | –0.5 | –0.125 | 0. | 0. |
| (1,2,4,7,5,1) | –0.5 | –0.1 | 0. | 0. |
| (2,4,8,7,5,2) | –0.5 | –0.1 | 0. | 0. |
| (1,2,4,8,7,5,1) | –1.0 | –0.1667 | 0. | 0. |
| (2,3,6,4,7,5,2) | 0. | 0. | 0. | 0. |
| (2,4,7,6,3,5,2) | 0. | 0. | 0. | 0. |
| (1,2,3,6,4,7,5,1) | –0 | –0.0714 | 0. | 0. |
| (1,2,4,7,6,3,5,1) | –0.5 | –0.0714 | 0. | 0. |
| (2,3,6,4,8,7,5,2) | –0.5 | –0.0714 | 0. | 0. |
| (2,4,8,7,6,3,5,2) | –0.5 | –0.0741 | 0. | 0. |
| (1,2,3,6,4,8,7,5,1) | –1.0 | –0.125 | 0. | 0. |
| (1,2,4,8,7,6,3,5,1) | –1.0 | –0.125 | 0. | 0. |
| *Mean, all cycles* | | –0.07682 | 0. | 0. |

As noted before, from (7)-(12) it is straightforward to calculate the number of points gained or lost around cycles. We see that the per-step performance of strategies averaged over all possible strategies is neutral around every cycle ($\Delta W = 0 = {}^{\Delta W}\!/_{\Delta t}$), while agent performance is $\sim -0.08$ averaged over all cycles. (The unit step change in wealth must be adjusted for the differing number of steps in different cycles. This has been done in columns 3 and 5 of **Table 8** )

Note, too, that the change in wealth for strategies when averaged over all possible cycles *and all* possible strategies is zero  For almost all given quenched disorder matrices the distribution of strategies is asymmetrical, and the asymmetry of this distribution, in conjunction with the minority rule for winning, ensures an average loss (even if under most circumstances less of a loss than for agents [2]). In other words, a "crowd" of strategies in one region of strategy space insures that their decisions will on average be the majority, hence losing decisions. By averaging over all possible strategies, we eliminate the expected asymmetry in strategy distribution.

Similar patterns are found for all but a few exceptional values of $\tau$ where mean agent performance is also neutral. (That this must be so follows from the fact that the mean performance of all agents in a given $\hat{\Omega}$ is at best 0.) Thus, over many different $\hat{\Omega}$, the optimization rule of the THMG degrades mean performance relative to the measured performance of underlying strategies. In Ref. [2], the inclusion of agents who select their worst-performing strategy is discussed. It turns out that these agents outperform not only standard agents, they outperform their underlying strategies and can even regularly attain net positive gain which for a standard agents in a MG structure is exceedingly rare. The performance of such agents around cycles have the same values as standard agents but with the opposite sign.

## 4.5 CYCLES, ANTIPERSISTENCE AND THE ILLUSION OF CONTROL

As discussed in [2], the degree to which agents underperform their own strategies varies with the phase as parameterized by $\alpha$ [17]. Our simulations have been performed in the "crowded" phase—the number of agents is significantly larger than the number of strategies in the RSS. Several agents are therefore likely to share strategies. As noted in [11], the "crowd" of such agents at any given time-step acts like a single "super-agent"; the remaining agents as a (non-synchronized) "anticrowd" whose actions will conform to the minority choice. Thus, when a strategy is used, it is probably used by more than one agent, often by many agents. If by enough, it becomes a losing strategy with large probability—precisely because so many agents "think" it's the best choice and use it. This implies that at the next time step, agents will not use it. The time-series of choices $\bar{A}_D$ therefore does not show trends (or persistence), but rather anti-*persistence*.

Anti-persistence is not equivalent to "random" and is scale-dependent. Consider a binary time-series with an $m$-bit $\mu(t)$ defined in the same way as we have in the MG or THMG: $\mu(t)$ is a window of length $m$ *sliding with a time step of one unit*: $\mu(t)=\{s(t-m+1),\ldots s(t)\}$ A *perfectly* anti-persistent binary series at scale $m=2$, for example, is characterized as follows: Select any one instance of the four possible $\mu(t)\in\{00,01,10,11\}$. Identify the following bit $s(t+1)\in\{0,1\}$. Now identify the next instance of the selected $\mu(t)$. If the series is perfectly anti-persistent, the following bit will be 1 if the previous following bit was 0, and 0 if the previous following bit was 1.

Such a series can be generated by two lookup tables indicating what bit follows which $\mu(t)$. Whatever bit is indicated by the first table, the opposite bit is indicated by the second. Whenever a entry in a table is used for a given $\mu(t)$, the other table is used when $\mu(t)$ occurs again [18]. These tables are identical to strategy pairs at the maximum Hamming distance in the MG. No matter which of the 16 possible strategies is used for table 1, and regardless of which of the 4 possible $\mu(t)$ are used to initiate it, the time series generated by these tables will rapidly settle into perfect anti-persistence.

Ignoring the initialization steps and steps to convergence, it turns out that such a perfectly anti-persistent series will always follow one of the two cycles on the binary deBruijn graph of order 3 that touch every possible 3-bit state: $\{1,2,4,8,7,6,3,5,1\}$ *et. cyc.* or $\{1,2,3,6,4,8,7,5,1\}$ *et. cyc.* Around either cycle the net mean gain averaged over all possible strategies is 0; but the net mean per-step change in wealth averaged over all possible strategy pairs (agents) is $-0.125$ (**Table 6**). Over the entire (unique 8-step) cycle, the net mean change over all strategy-pairs is exactly $-1.0$. (Indeed, over 8 steps, the net mean change in wealth for all cycles is either $-1.0$ or $0$.)

By definition, a cycle consists of sequence of states attained only once except for the return to the initial state. For memory of length $m$ ($m+\tau$ in the case of the THMG), the number of states is the longest possible cycle $L_{\max}\equiv 2^m+1$ (or $2^{m+\tau}+1$ in the THMG). In any cycle composed of fewer states than $L_{\max}$, the return to the initial state will therefore be persistent. The transition probabilities in a general $\hat{\mathbf{T}}$ representing $000 \rightarrow 000$ and $111 \rightarrow 111$, which are in principle allowed, in the THMG are always zero. (I.e., $000 \rightarrow 000$ and $111 \rightarrow 111$ are disallowed for $m + \tau = 3$; $0000 \rightarrow 0000$ and $1111 \rightarrow 1111$ are disallowed for $m + \tau = 4$, etc.) This follows from the fact that these transitions—which are the same as the two *minimum* length cycles—are perfectly persistent (at the minimum $m'$).

Either $\{1, 2, 4, 8, 7, 6, 3, 5, 1\}$ *et. cyc.* or $\{1, 2, 3, 6, 4, 8, 7, 5, 1\}$ *et. cyc.* could hypothetically be generated by a crowd of agents in the MG or THMG all sharing strategies that lead to identical actions (and a correspondingly fixed "anti-crowd" leading to opposite actions) at every time step. In the MG or THMG this could only occur perfectly if every agent shared the same pair of strategies at the maximum Hamming distance. To the extent that agents are diverse and with a variety of Hamming distances, the resulting time series is less than perfectly anti-persistent. A "superagent" or crowd forms on the basis of good prior performance by strategies: Its action for a given $\mu(t)$ is selected by many agents. But the minority rules ensures that its wide selection makes it more likely to lose than to win. The strategy not being used will be a "minority" strategy which has more chance to win (as formally confirmed in [19]). Furthermore, formation of a cohesive "superagent" or crowd requires a high probability that many agents will select the same action at a given time. This occurs in the so-called "crowded" phase when there are "many" agents relative to the number of strategies.

Thus, the illusion of control effect is fundamentally due to three ingredients: (i) the minority mechanism (an agent or a strategy gains when in the minority and loses otherwise); (ii) the selection of strategies by many agents because they were previously in the minority, hence less likely to be so in the present; and (iii) the crowding of strategies (i.e. few strategies for many agents). In many social and economic activities, we humans as agents attempt to maximize value. We often do so by adjusting our present strategy in the light what of has previously worked best for the "winners". Yet this very adjustment often proves to have exactly the opposite effect—causing greater losses than if we had left well enough alone. A classic everyday example which has been analyzed in these terms is weaving in and out of traffic—we rarely gain, and often lose by doing so. We would do better sticking to whatever lane we find ourselves in [20]. The negative power of this effect is demonstrated by the perverse phenomenon which we have here mentioned as well (and discussed in detail in [2]): That in certain games, deliberately selecting what appears to be the worst strategy can "paradoxically" enhance gains.

# 5 . CYCLE DECOMPOSITION OF MG, MAJG AND $G AND FINANCIAL TIME SERIES

The MG has been widely discussed as a model market with $A(t)$ (or $Sgn\left[A(t)\right]$ ) equivalent to a price time series (or as the series of signs of change in price). A mechanism for the appearance of pockets of predictability in the MG has been demonstrated in [21] and successfully applied to the prediction of a real-world financial time series, the NASDAQ composite index (symbol IXIC). In the case of the THMG, the appearance of such pockets follows directly from complementary entries of 1 and 0 in $\hat{\mathbf{T}}$, as for example in eqn. (42). This will happen following any $\mu_t$ for which the values of $A_D\left(\mu_t\right)$ is sufficiently large in conjunction with $N_D\left(\mu_t\right)$ such that no possible value of $A_U\left(\mu_t\right)$ (in conjunction with $N_U\left(\mu_t\right)$) can reverse $Sgn\left[A(t)\right]$. Ref. [21] shows that such deterministic sequences occur with greater frequency than one might intuitively expect and demonstrates that the IXIC composite index (IXIC) similarly shows a surprisingly high frequency of such pockets. From the perspective of cycles, we likewise anticipate that any time-series whose cycle decomposition demonstrates a significant weighting of certain cycles than would be expected by chance should have a higher degree of determinism and possibly, too, predictability. We demonstrate this phenomenon using the IXIC composite index decomposed into cycles at m=2. Based on rank-ordered cycles, we create a toy predictor that forecasts change in direction one step ahead.

## 5.1 CYCLE DECOMPOSITION OF MG, MAJG and $G

In the MG/THMG, MAJG/THMAJ and $G/TH$G, agents attempt to predict the next time-step in the "market" in which they are participating: The market's "value", i.e. A(t), is simply the sum of all agents' predictions. It has long been known that in the MG/THMG, agents cannot *in general* succeed more often than not—assuming no privileging, i.e., all agents have the same memory and access to strategies.

More precisely, in the regime $m > m_c$, i.e., where the time series is modestly persistent and approaching the random limit with increasing $m$, a majority of agents will lose persistently but a small subset will win persistently. However, *which* agents are winning (i.e., which strategies they have) depends upon the particular configuration (strategy composition) of all other agents. Since agents do not know what the initial quenched disorder is, agents cannot intelligently choose the ideal strategies for a given such disorder. And if all agents could do so, they would thereby reconfigure the initial disorder in a potentially infinite recursion to no avail. Likewise, as discussed in Ref.[2], when agents are allowed to vary their strategy composition in real-time based on past performance, the system as a whole converges to the performance limit of a set of random or fixed choice agents. Thus, even in a regime where some agents do consistently win, their winning is a matter of luck, i.e., it follows from the strategies they happen to be assigned vis-à-vis the strategies randomly assigned to all other agents.

On the other hand, in the regime $m \leq m_c$, all agents behave much like random or fixed choice agents. That is, no agent can consistently win. Put differently, for $m > m_c$ there is information in the history $\mu(t)$ (of length $m$) in the restricted sense that there is something that at least some agents can use, by their good fortune, to make predictions. For $m \leq m_c$ there is no such information to be extracted by any agent.

However, a privileged agent with access to a longer history than others will in general be able to outperform non-privileged agents at all $m$. Likewise, an outside non-participant with access to an arbitrary amount of past history should be able to predict future steps of the time series with a better than chance success rate. From this perspective there exists information in the time-series at all $m$. The nature of this information can be highlighted by looking at the cycle structure of the time-series and can be tested by implementing predictors based on this structure.

In **Figure 4** we presented a side-by-side illustration of the respective cycle structure for the MG, MAJG and \$G at a single value of $m'$, with the raw cycle weights transformed into an excess or deficiency relative to that expected in the decomposition of a random binary series. The most striking feature of this representation is perhaps not so much their differences from each other as their differences from the decomposition of the NASDAQ composite index (IXIC): All three games generate series with cycle decompositions that are roughly the inverse of the IXIC series decomposition.

In **Figure 11** we present a side-by-side comparison of the relative weights for the MG, MAJG and \$G at $2 \le m \le 8$ and $m' = 2$. The phase transition in the MG is evident in the change in the cycle structure at $m = 4$. It is important to note that these are ensemble averages over many different initial quenched disorder states. The cycle distribution for any single initial disorder usually differs substantially from any other and typically departs widely from that for a random series. Thus, for the MG, the closeness of the distribution to the random distribution, especially for small $m$, represents not the typical distribution but the many widely varying distributions around a mean.

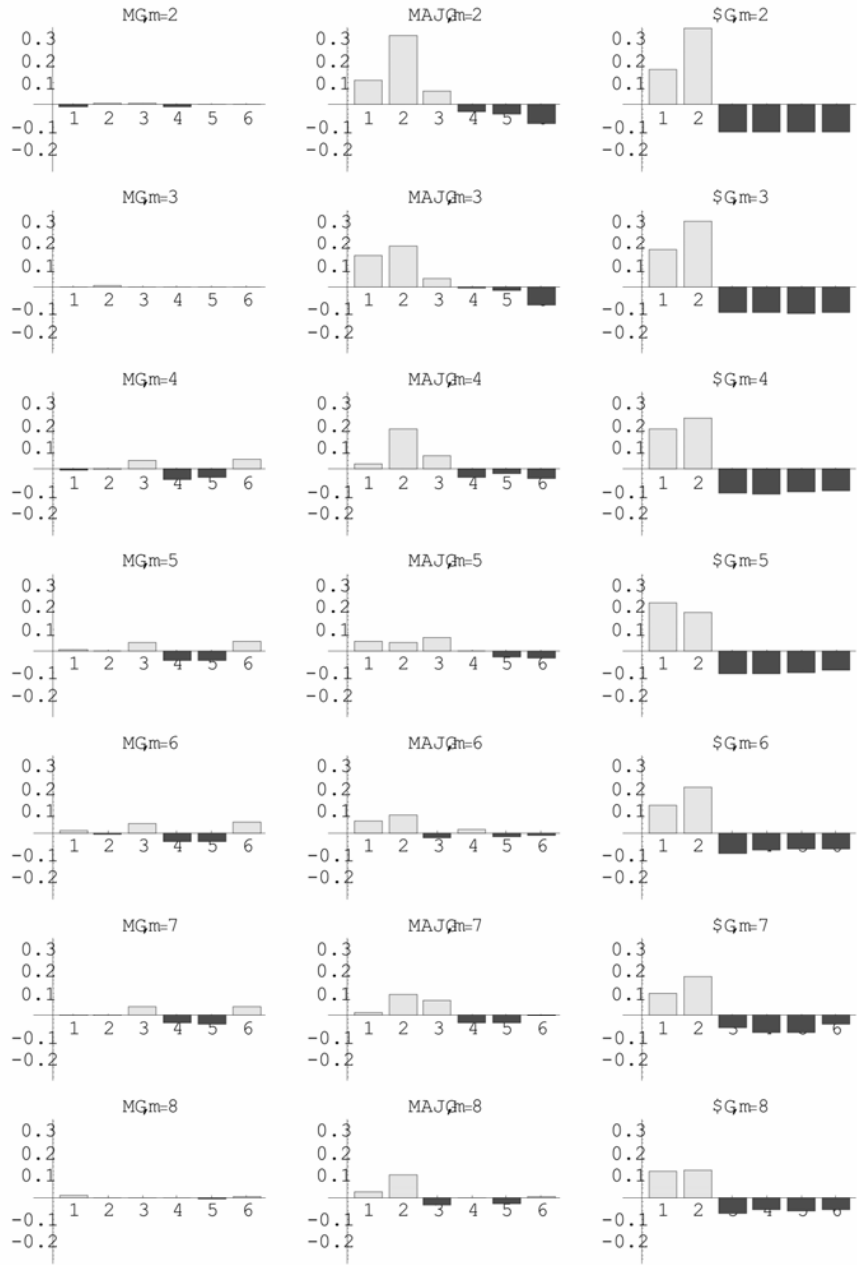

**Figure 11**: Relative cycle weights given $\{m, S, N\} = \{2, 2, 31\}$ for MG, MAJG and \$G from left to right and $2 < m < 8$ from top to bottom. Within each graph, the numbers 1-6 on the *x*-axis represents the six cycles at *m'*=2, i.e., $(1,2,3,4,5,6) \equiv \left( \{1,1\}, \{4,4\}, \{2,3,2\}, \{1,2,3,1\}, \{2,4,3,2\}, \{1,2,4,3,1\} \right)$. The cycles may likewise be converted into the outcome + or − following a sequence of directional changes per **Table 1**. I.e., if the last number in the cycle label is odd, the height of the bar represents the probability relative to a random series that − follows the series of directional changes; if even that a + follows.

## 5.2 DECOMPOSITION OF THE NASDAQ COMPOSITE INDEX

We first convert the daily values of the IXIC composite index (IXIC) into a binary series of price changes, 1 for up, 0 for down. If the price remains unchanged on two successive days, 1 or 0 is entered at random. (Note that this method is much simpler if less precise than a trinary series that excludes a necessarily large number of small price changes. A decomposition into base three cycles proceeds analogously to a decomposition into base two.) For m=1 the possible cycles that may arise are (1,1), (2,2) and (1,2,1) et cyc.. For m=2, (1,1), (4,4), (2,3,2) et cyc., (1,2,3,1) et cyc., (2,4,3,2) et cyc. and (1,2,4,3,1) et cyc. For m=3 there are 19 (et cyc.) unique cycles. States are denoted by 1 plus the decimal value of the binary sequence of length m. The cycle structure is determined by the method in section III.A. We know from section II.B. that up to a finite size error the result is equivalent to an analytic decomposition of a Markov-chain representation of the series treated as stationary. (For the MG, a stationary state is always eventually attained very rapidly in the THMG, for small m, $\tau$ ). The error decreases asymptotically with the number of binary data points used. The IXIC in its entirety is evidently non-stationary. Furthermore, the bubble and crash of the years ~1998-2002 depart from a simple linear correction on a visible scale. Thus the larger number of points included to create the decomposition the greater the accuracy with respect to finite-size error, but the less locally representative it will be. We choose data length equal to one trading year (250 trading days).

We first note that decomposing a time series into cycles—even one as simplified as this—provides an immediate impression of structural departures from randomness. For the 8700 trading days of the IXIC prior to June 1, 2007, the fraction of up days = 0.56. If we randomize a large number of 1's and 0's in respective proportions 0.56 and 1-0.56=0.44, the expected distribution of cycles for m'=2 is shown in **Table 7** alongside the actual distribution of cycles for the IXIC from this time period. In terms of persistence and anti-persistence, the actual IXIC series shows an excess of persistent cycles, i.e., (1,1) and (4,4), and a deficit of anti-persistent ones, (2,3,2) in particular. Recent research has shown that it is possible to devise working predictors by generating synthetic time series from MGs with a variety of parameters such that they are tuned to the characteristics of the real-world time-series targeted for prediction[22]. When a real series is tuned to a mix of both MG and Majority Games, better results follow than when only MG's are used, since the latter captures primarily anti-persistence at reasonable values of m while the Majority Game captures persistence[23].

**Table 9**. m'=2 cycle structure for IXIC binary series and IXIC-biased random binary series

| Cycle m'=2 | Frequency (IXIC-Biased) Random Series | Frequency IXIC Series | Frequency Difference IXIC-IXIC-Biased Random |
|---|---|---|---|
| (1,1) | 0.16 | 0.21 | +0.05 |
| (4,4) | 0.35 | 0.40 | +0.05 |
| (2,3,2) | 0.12 | 0.06 | -0.06 |
| (1,2,3,1) | 0.09 | 0.08 | -0.01 |
| (2,4,3,2) | 0.15 | 0.12 | -0.03 |
| (1,2,4,3,1) | 0.12 | 0.12 | 0 |

In column 3 of **Table 9**, we see the actual cycle weightings for the IXIC and in column 2 the actual weightings for a random binary sequence with the same upward bias as the IXIC. Column 4 provides the difference, thus excluding the net upward drift. Per **Figure 5** and the distances defined by equation (1), we have $d_{maj} = 0.270 > d_{\$} = 0.133 > d_{IXIC} = 0.118 > d_{min} = 0.028$. By this measure, the IXIC is less random than the MG series but more random than the MAJG and \$G series.

### 5.3  DECOMPOSITION OF THE PHILA GOLD AND SILVER INDEX

By contrast, **Figure 12** presents the relatively-weighted cycle decomposition for the entire history of the XAU (Philadelphia Gold and Silver Index; December, 1983 through February, 2008) with $d_{XAU} = 0.057$—approximately twice as close to random as the IXIC. The XAU shows 49% up days with an annualized buy and hold return of 0.030. A cycle predictor based on this decomposition predicts 51% of the daily directional changes correctly but yields an annualized return of only .019. Unsurprisingly then, with a series much closer to a random one by both the cycle decomposition method and the measure of persistence, prediction by the cycle decomposition method is much more difficult and in fact fails.

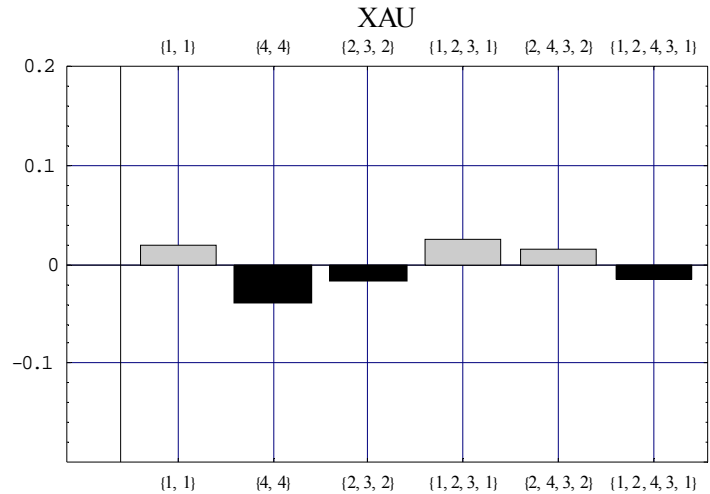

**Figure 12**: Relative-weight cycle decomposition of the Philadelphia Exchange Gold and Silver Index (XAU)

## 5.4 CONSTRUCTING THE PREDICTOR(S)

The cycle decomposition predictor itself consists of a table regenerated from all $m'$=2 cycles for every possible 250-day (-step) sliding window in sequence along a binary series. The cycles are rank-ordered by weight (frequency). The last state prior to the prediction day consists of 1 plus the last $m$ days of binary data converted to decimal form. The predictor simply consists of the next state cycle following the present state. The prediction is the first number (1 or 0) in the binary representation of the state. If the present state is represented in more than one cycle, the cycle with the larger weight is used, consistent with a $0^{th}$-order approximation by a wholly deterministic process (The rare persisting ties are settled by a fair coin toss. The dependency predictor is constructed similarly. The only procedural difference is that decimally-converted last state prior to the prediction date identifies a unique member from the set of dependencies, hence generates a unique prediction.)

We start using days 1-250 predicting day 251 and then slide the 250 day window forward by one day and recalculate the decomposition (and dependencies), so that days 2-251 predict day 252. The process is repeated through all days of data (minus the initial 250) and the percentage of correct predictions is calculated. These results are compared to a "buy and hold" strategy which is equivalent to always predicting 1.

We find that even for so simple a cycle predictor the fraction of its correct predictions is ~0.58 which on the face of it appears significantly better than what may attained by chance and better than a "prediction" consisting of "always assume the next day's change will be positive", i.e. a "buy and hold" strategy.

One might naïvely use the t-test and the binomial distribution to calculate p ≤ 0.00008 for the 0.58 fraction of correct predictions. But this value is not likely to be accurate for the same reason that **Table 9** demonstrates such large departures from a random structure, hence Gaussian distribution: The underlying distribution of up and down days results from short-term (a few days of ) internal dependencies as discussed for example in Ref. [10] with respect to the Dow Jones Industrial Average. We use these dependencies to obtain a more realistic and empirical "non-Gaussian p" that reflects the exact distribution of the one real IXIC time-series available to us.

The method we employ is as follows. First, consider what is required to obtain from simulations a (Gaussian) p-value for 0.58 correct predictions: Flip a 0.56 heads-biased coin 8700 times, many times over, to generate a large number of 8700-bit synthetic sequences (Heads = 1, tails = 0.). These sequences are memoryless (m= 0): each bit is independent of any other. Then count the number of such sequences for which the mean is ≥ 0.58 and divide by 8700. This reproduces the p-value generated by the binomial distribution.

Second consider that the real series shows tendencies toward persistence or anti-persistence (perhaps both in different regions and even if towards persistence overall—as the 0.56 mean would imply). Then these dependencies will tend to perpetuate themselves with some probability (thus tending to perpetuate even further). This yields fat-tailed distributions that are similar to financial market returns where extreme events are more likely than under the Gaussian assumption of complete independence of price changes. It likewise causes more than the expected number of runs of certain shorter lengths (two- or three-bit states, say, and beyond the overall 0.56 + bias) and fewer of others. (See Ref. [24] for detailed studies of this phenomenon in terms of market "drawdowns".)

We characterize such dependencies at the *m*-bit level for a range of *m* as follows. *m*=0 is defined as no dependency: There is simply a 0.56 probability at every time-step that the next bit will = 1 (and a 0.44 probability that it will = 0). To determine the *m*=1 dependency, determine the proportion of instances (out of a total of 8700-1) in which 0 is followed by 1 and in which 1 is followed by 1. (This is equivalent to recasting the series in paths of length *m*+1=2, converting to digital representation $\{0,1,2,3\}$ and finding the proportion of the two odd values, 1 and 3: $0 \rightarrow 1 \equiv 01 = 1$ and $1 \rightarrow 1 \equiv 11 = 3$) Under a Gaussian assumption, the frequency (probability) of either odd value (relative to their even alternatives) would simply be 0.56. But in fact the probability of a 1 appearing is dependent on the value of the preceding bit: $p(1|0)$=0.48; $p(1|1)$=0.63. This is the simplest demonstration of persistence in the IXIC. Over 8699 days the departure from the Gaussian expectation is significant at greater than 6 standard deviations as shown in **Figure 13**.

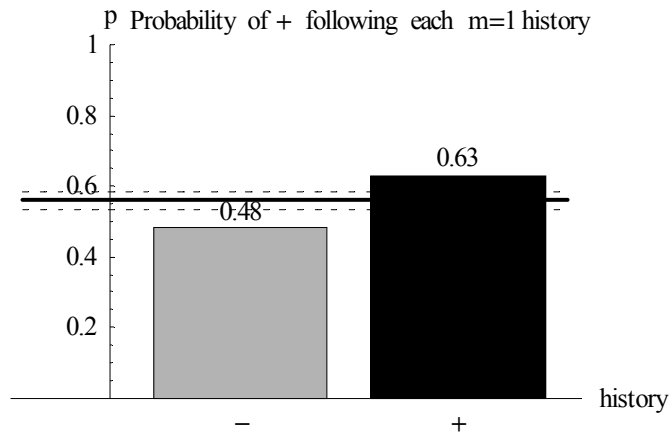

**Figure 13**: The probabilities that an up day will follow a down day (–) or an up day (+) in the IXIC based on nearly the entire daily history (8700 days) of price changes. The solid black line at 0.56 represents the overall proportion of + price changes hence the Gaussian expectation of +. The two dashed lines above and below represent departures of 3 standard deviations from 0.56.

A similar analysis based on the four possible two-day prior histories is shown in **Figure 14**. All four histories ($\{00,01,10,11\} \equiv \{--,-+,+-,++\}$) are associated with succeeding probabilities for 1 (+) that again depart from the Gaussian expectation of 0.56 at a significance of greater than 3 standard deviations. (This analysis is similar to that performed by Zhang in [10], but we use it here not to argue against the efficient market hypothesis but as a practical means of assessing non-Gaussian probabilities.)

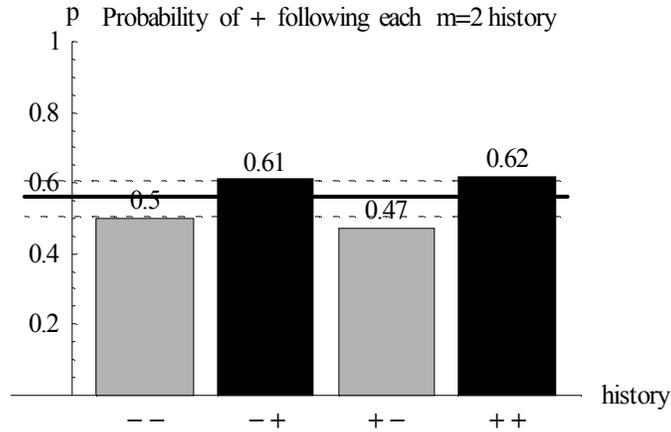

**Figure 14**: The probabilities that an up day will follow each of $(--),(-+),(+-)$ or $(++)$ in the IXIC based on nearly the entire daily history (8700 days) of price changes. The solid black line at 0.56 represents the overall proportion of + price changes, hence the Gaussian expectation of +. The two dashed lines above and below represent departures of 3 standard deviations from 0.56.

As longer histories are used, the individual probabilities in general depart from the Gaussian expectation at lower significance levels and the alternation between <0.56 and >0.56 based on the immediate prior bit attenuates greatly. Most of the simple dependency structure appears to be captured by the probabilities that follow a two-bit history.

With these empirically-determined dependencies create a large number of synthetic, non-Gaussian time series in the following way: Generate a random two-bit initial state and generate the succeeding +1 or −1 with the properly associated probability, i.e. $p(1|00) = 0.501, p(1|01) = 0.611, p(1|10) = 0.468, p(1|11) = 0.621$. Shift forward 1 step and repeat 8698 times. As a result, the appropriate degree of persistence is captured in each replica of the original. Many more such replicas show a 0.58 net +1 imbalance or greater than in a Gaussian simulation while the overall expectation value for 1 remains 0.56.

From these replicas we may compute a more realistic *p*-value for the toy predictor (which was likewise based on *m'*=2). In this case, we obtain p ≤ 0.00014 based on 10,000 replicas of the 8700 day series.

# 6. DISCUSSION AND FUTURE DIRECTIONS

We have presented two tools for analyzing binary series—persistence and the cycle decomposition and for generating predictions of future binary data points based on prior data. Both methods arise as a natural means for teasing apart the statistical structure of the time-series generated by MG, MAJG and $G all of which were designed to capture certain aspects of real-world markets. When the underlying structure of a system generating a time-series is known, and especially when it is known to be Markovian, the cycle decomposition is analytically exact and provides an explicit portrait of the type and degree of determinism embodied in the system. The microscopic performance of agents and their strategies may be studied by examining their deterministic performance around each cycle. Even peculiar effects such as the illusion of control in MG—an illusion found commonly in many real world attempt to control and predict—can be understood in terms of the different kinds of behavior displayed as a system traverses different cycles.

As we have shown, it is possible to generate predictors based on the cycle decomposition which under certain circumstances yields efficient real-world results, even when the predictors are extremely simple, even naïve. One intuitively sensible such circumstance is that the series being predicted departs from randomness in a way that he predictors can exploit. Indeed, both the cycle decomposition method and the calculation of the persistence of a series provide measures of to what extent a series does depart from randomness.

Both the measure of persistence and the cycle decomposition method require the selection of a memory length $m'$ at which to perform the analyses. We do not here report on the results of the next evident step, namely the analysis of empirical series at multiple $m'$ (since the natural $m'$ is unknown or may not exist). However, as an example, it is generally true that series that are predictable using these methods show consistently non-random values for persistence at multiple $m'$, while unpredictable ones do.

For example, as shown in **Figure 15**, even when the IXIC is made linearly stationary (to eliminate the upward bias), the persistence parameter $\vartheta_{IXIC}$ shows a significant deviation from the expected value 0.5 for a random series at all values $1 < m' < 10$, and with a mean over these ten values of $0.52 \pm 0.004$. By contrast, the linearly stationary $\mathcal{P}_{XAU}$ shows no such deviation from the random expectation, with a mean over the same range of exactly $0.50 \pm 0.008$.

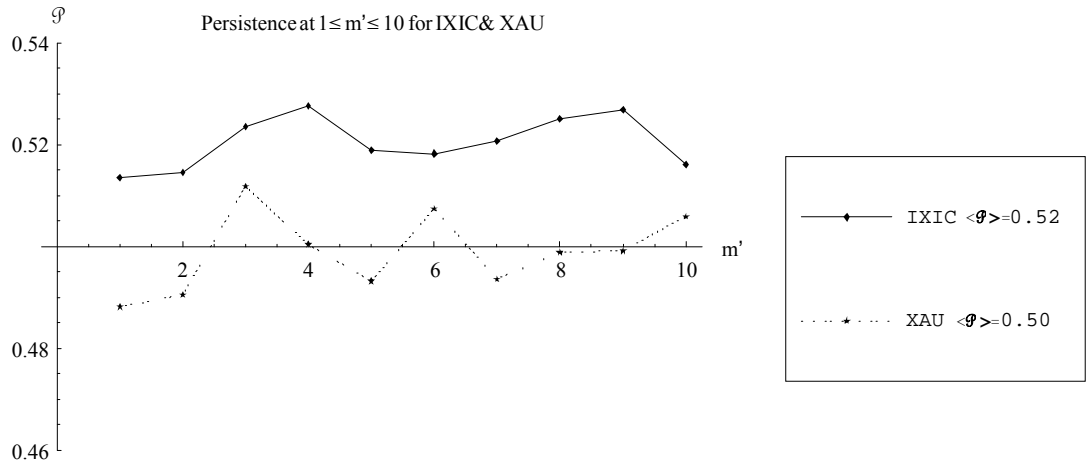

**Figure 15**: Persistence measure $\mathcal{P}$ for the IXIC and XAU at for values of $1 \le m' \le 10$. (Recall that $\mathcal{P}$ is the fraction of binary sequences of length m' within a series which if followed by 1 at a given appearance, will be followed by 1 again at its next appearance, averaged over all appearances and all possible sequences of that same length.)

Many of the results pertaining to real-world time-series in this paper relate to the extensive literature on "technical analysis" to predict financial time series. Technical analysis may be defined as methods of predicting price changes or change of direction based chiefly on prior prices, sometimes also on open interest an/or on volume. For many years, technical analysis was exclusively the domain of amateur and professional traders and academic proponents of the efficient market hypothesis (EMH) presumed that prediction on such a basis was a priori impossible. More recent statistical studies have demonstrated both that certain forms of technical analysis (especially those that avoid the dangers of excess data-mining) may indeed be profitable; and that the profitability of any given method has declined over time, especially since the advent of rigorous computer-based quantitative algorithms of ever-increasing speed and sophistication. Both points are reviewed in [25], while [26] demonstrates how technical analysis has incorporated statistical inference with the largely intuitive pattern-recognition methods of twenty and more years ago. As more methods of increasing complexity are introduced into a market, that market becomes progressively more efficient. The remaining marginal inefficiency therefore becomes that much more difficult to exploit, and can readily reach the point where (for an indefinite period of time, if not permanently) the costs associated with extracting any residual potential gain remain greater than the gain itself. It is also the case that not all markets are equally non-exploitable. Thus, for many years, the Standard and Poor' Index of 500 American stocks as well as the precious metal markets have been considered exemplars of efficiency beyond practical exploitation by technical methods. By contrast, staple commodity markets have until very recently largely remained the province of old-fashioned floor traders with a significant degree of inefficiency to be exploited. $\vartheta$, the measure of persistence we apply to real-world data may provide a simple measure of how likely it is that technical measures will succeed, or put more precisely, how slim is the marginal inefficiency. The contrast in $\vartheta$ shown in **Figure 15** suggests that the precious metals market is highly efficient but that(until more recently) the IXIC has not been.

The decline in profitability over time of a given method is strikingly demonstrated in our own discussion of a cycle-decomposition predictor (see **Figure 9**) as is another point: That when a given method's returns decline to the level where cost exceeds benefit, other more sophisticated improvements may (but surely only for a time) restore profitability.

Many prior studies have indicated the presence of linear price-change correlations in high-frequency financial data e.g., ref.s [27][28][29][30], but not in low-frequency data (hourly, daily, weekly, etc.) The presence of linear correlations suggests that persistence ought to be in general higher in higher-frequency data than lower. **Figure 16** illustrates this for the US Dollar/Deutsche Mark exchange rate for October 1, 1992 (made linearly stationary over the day). It demonstrates a mean $\vartheta_{USD/DM} = 0.55$, much higher than for the IXIC and very far from random—consistent with the observation that foreign exchange markets remain susceptible to technical trading methods [25]. Note that for m' = 10, patterns of up to $2^{10} = 2048$ binary bits show a significantly greater tendency to repeat than to not, implying a degree of non-linear and multi-step correlation above and beyond any linear correlations.

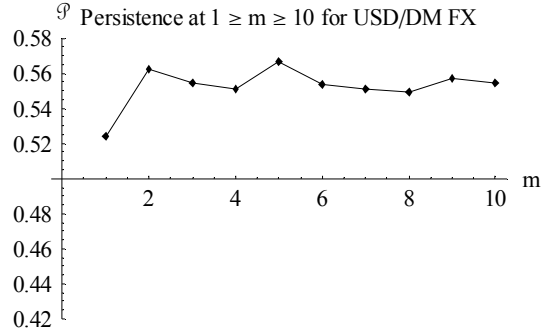

**Figure 16**: Persistence at m between 1 and 10 for high-frequency (tick) data for the USD/DM foreign exchange rate on Oct. 1, 1992 (~6500 data points)

Similarly, the cycle decomposition shows $d_{USD/DM} = 0.27$, equivalent to $d_{MAJ}$. If one imagines it were possible to trade without cost long or short on any tick, an $m' = 2$ cycle decomposition predictor would yield 61% correct choices of direction and a one-day gain of 1.61 versus 58% up-ticks with a buy and hold gain of .005.,A one-day gain of 1.61 translates into an annualized return of $> 10^{51}$)—the excessively large number reflecting the absence of the enormous trading and other costs associated with such rapid theoretical trading). Down-ticks are in general much larger than up-ticks; the cycle decomposition predictor tacitly accounts for this discrepancy in efficient fashion such that the ~3 percentage point improvement in directional prediction translates to a very much more impressive gain.

Similar consistency with respect to the configuration of the respective cycle decompositions and pseudo-linear distances from random are found across many high-frequency and daily price indices. This fact is consistent with the ability of a persistence filter to enhance the efficiency of a cycle-decomposition predictor as discussed above. It would be valuable to conduct a more wide-ranging survey of various time-series, both real-world and synthetic to quantify the reliability of these measures and the predictors described.

We have here concentrated upon cycles and persistence as defined for binary series. While computationally more intense, it is straightforward mathematically to extend these measures to trinary, quaternary, etc. series. Preliminary studies demonstrate greatly improved predictive power, for example, when parsing real-world price changes into three rather than two domains (i.e., $\{+,0,-\}$ instead of $\{+,-\}$) and thus generating trinary series and the associated measure of persistence.

We look forward to extending and reporting on the application range of these relatively simple and intuitive measures, for analyzing and characterizing time-series as a whole, for characterizing different regimes within time-series, and for generating prediction methods.

## REFERENCES

[1] M. L. Hart, P. Jefferies, and N. F. Johnson, Physica A **311** (1), 275 (2002).

[2] J. B. Satinover and D. Sornette, Eur. Phys. J. B **60** (3), 369 (2007).

[3] J. B. Satinover and D. Sornette, Illusory versus genuine control in agent-based games, submitted to Eur. Phys. J. B. (2008).

[4] P. Jefferies, M. L. Hart, and N. F. Johnson, Phys. Rev. E. **65** (1 Pt 2), 016105 (2002).

[5] P. Cvitanovic, Chaos: An Interdisciplinary Journal of Nonlinear Science **2** (1), 1 (1992).

[6] B. Eckhardt, Periodic orbit theory, Proceedings of the International School of Physics "Enrico Fermi" 1991, Course CXIX "Quantum Chaos", G. Casati, I. Guarneri, and U. Smilansky (eds.), pp. 77-118, North Holland (1993).

[7] T. Ryden, T. Terasvirta, and S. Asbrink, Journal of Applied Econometrics **13** (3), 217 (1998).

[8] B. Knab, A. Schliep, B. Steckemetz, and B. Wichern, Between Data Science and Applied Data Analysis, M. Schader, W. Gaul, and M. Vichi, eds., Springer, 561 (2003).

[9] C. P. Papageorgiou, presented at the Computational Intelligence for Financial Engineering (CIFEr), 1997 (unpublished).

[10] Y. C. Zhang, Physica. A **269** (1), 30 (1999).

[11] M. Hart, P. Jefferies, P. M. Hui, and N. F. Johnson, The Eur. Phys. J. B **20** (4), 547 (2001).

[12] D. Challet and Y. C. Zhang, Physica A **246** (3), 407 (1997); D. Challet and Y.-C. Zhang, Physica A **256** (3-4), 514 (1998).

[13] S. Kalpazidou and J. E. Cohen, Circuits, systems, and signal processing **16** (3), 363 (1997).

[14] Q. Minping and Q. Min, Probability Theory and Related Fields **59** (2), 203 (1982).

[15] S. E. Kalpazidou and N. E. Kassimatis, Circuits, Systems, and Signal Processing **17** (5), 637 (1998).

[16] S. Kalpazidou, Journal of Applied Probability **29** (2), 374 (1992).

[17] D. Challet, A. D. Martino, M. Marsili, and I. P. Castillo, Journal of Statistical Mechanics: Theory and Experiment **2006** (03), P03004 (2006).

[18] R. Metzler, J. Phys. A **35** (25), 721 (2002).

[19] R. D'Hulst and G. J. Rodgers, Physica A **278** (3), 579 (2000).

[20] T. Chmura and T. Pitz, Physica A **363** (2), 477 (2006).

[21] J. V. Andersen and D. Sornette, Europhysics Letters **70** (5), 697 (2005).

[22] N. F. Johnson, D. Lamper, P. Jefferies, M. L. Hart, and S. Howison, Physica A **299** (1-2), 222 (2001).

[23] C. Gou, Neural Networks and Brain, 2005. ICNN&B'05. International Conference on **3** (2005).

[24] D. Sornette, *Why Stock Markets Crash: Critical Events in Complex Financial Systems*. (Princeton University Press, 2002); A. Johansen and D. Sornette, The Eur. Phys. J. B **1** (2), 141 (1998); A. Johansen and D. Sornette, Journal of Risk **4** (2), 69.

[25] C. H. Park and S. H. Irwin, Journal of Economic Surveys **21** (4), 786 (2007).

[26] D. R. Aronson, *Evidence-based technical analysis: applying the scientific method and statistical inference to trading signals*. (Wiley, New York, 2007).

[27] C. P. Papageorgiou and C. P. Papageorgiou, presented at the Computational Intelligence for Financial Engineering (CIFEr), 1997., Proceedings of the IEEE/IAFE 1997, 1997 (unpublished).

[28] G. Bonanno, F. Lillo, and R. N. Mantegna, Quantitative Finance **1** (1), 96 (2001).

[29] M. Lundin, M. M. Dacorogna, and U. A. Müller, Financial Markets Tick by Tick, 91 (1999).

[30] B. Zhou, Journal of Business & Economic Statistics **14** (1), 45 (1996).